\documentclass{aa}

\usepackage[varg]{txfonts}
\usepackage{amsmath,bm}
\usepackage{booktabs}
\usepackage{graphicx}
\usepackage{hyperref}
\usepackage{xcolor}
\usepackage{siunitx}
\usepackage{tabularx}
\usepackage{array}
\usepackage{placeins}
\usepackage{multicol}
\usepackage{longtable}

\hypersetup{
  colorlinks=true,
  citecolor=blue!55!black,
  linkcolor=blue!55!black,
  urlcolor=blue!55!black
}
\newcommand{\nsat}{n_{\mathrm{sat}}}
\newcommand{\Esat}{E_{\mathrm{sat}}}
\newcommand{\Ksat}{K_{\mathrm{sat}}}
\newcommand{\Qsat}{Q_{\mathrm{sat}}}
\newcommand{\Zsat}{Z_{\mathrm{sat}}}
\newcommand{\Jsym}{J_{\mathrm{sym}}}
\newcommand{\Lsym}{L_{\mathrm{sym}}}
\newcommand{\Ksym}{K_{\mathrm{sym}}}
\newcommand{\Qsym}{Q_{\mathrm{sym}}}
\newcommand{\Zsym}{Z_{\mathrm{sym}}}
\newcommand{\Mmax}{M_{\mathrm{max}}}
\newcommand{\Msun}{M_{\odot}}
\newcommand{\dd}{\mathrm{d}}
\newenvironment{ruledtabular}{}{}
\newcommand{\colrule}{\midrule}

\newcommand{\chEFT}{\ensuremath{\chi\mathrm{EFT}}}

\makeatletter
\newenvironment{appendixtable}{\par\penalty0\addvspace{12pt}\noindent\begin{minipage}{\textwidth}\def\@captype{table}}{\end{minipage}\par}
\makeatother
\begin{document}
\nolinenumbers
\renewcommand{\figurename}{Figure}

\title{What constrains the neutron-star crust?}
\subtitle{Present data, future measurements, and EOS doppelg\"angers}
\titlerunning{What constrains the neutron-star crust?}

\author{Vishal Parmar\inst{1,2,3}\corrauth{vishal.parmar@fair-center.eu}}
\authorrunning{Vishal Parmar}
\institute{Facility for Antiproton and Ion Research in Europe GmbH (FAIR),
Planckstra\ss e 1, 64291 Darmstadt, Germany
\and
GSI Helmholtzzentrum f\"ur Schwerionenforschung GmbH,
Planckstra\ss e 1, 64291 Darmstadt, Germany
\and
Istituto Nazionale di Fisica Nucleare (INFN), Sezione di Pisa,
Largo Bruno Pontecorvo 3, I-56127 Pisa, Italy}

\date{Received --; accepted --}

\abstract{The neutron-star crust plays a central role in pulsar glitches, thermal evolution, and stellar oscillations, yet its properties remain difficult to constrain. We use Bayesian inference with a unified crust--core equation of state to assess what present nuclear and astrophysical data tell us about the crust and what future stellar measurements can add. Within this model, present crust constraints are dominated by chiral effective field theory, with laboratory data providing additional information and current stellar observations having a smaller effect. The adopted theoretical bands determine how tightly the crust properties are constrained, while the treatment of neutron-rich surface energy affects their inferred values. Forecasts centred on a single reference equation of state show that more precise measurements of radius, tidal deformability, and total moment of inertia mainly constrain crust thickness. They provide little additional information about the crust--core transition, crust mass, or crustal moment of inertia and do not recover the precision lost when the chiral effective field theory constraint is removed. Two models with nearly identical global stellar observables can have crust masses and crustal moments of inertia differing by about a factor of two, yet remain indistinguishable at the assumed precision. Precise global measurements therefore need not uniquely determine the crust. Observations that directly probe the crust are needed to distinguish these models.}

\keywords{stars: neutron -- dense matter -- equation of state --
methods: statistical -- gravitational waves}

\maketitle
\nolinenumbers
\let\linenumbers\nolinenumbers

\section{Introduction}
\label{sec:introduction}
The neutron-star (NS) crust contains only a small fraction of the stellar mass
and radius, yet it substantially influences a wide range of observational
phenomena.  Its properties are central to models of pulsar glitches
\citep{AndersonItoh1975,Link1999,Andersson2012Glitches,Chamel2013Entrainment}, 
cooling of transiently accreting stars \citep{BrownCumming2009,PageReddy2013}, 
magnetar oscillations and crust failure
\citep{Israel2005QPO,StrohmayerWatts2006,HorowitzKadau2009,ParmarBombaci2026QPO}, pre-merger crust
shattering, and the ejection of neutron-rich matter
\citep{Tsang2012Shattering,Goriely2011CrustRProcess}.  Thermal and magnetic
transport through the crust also helps set the observed X-ray surface temperature
\citep{ChamelHaensel2008,Geppert2004Temperature}. Despite its importance for NS observables,
the composition, microscopic structure, and extent of the crust remain difficult to constrain.
Although laboratory nuclei constrain matter at sub-saturation densities,
the extreme neutron excess, nuclear lattice, and unbound neutrons of the
inner crust remain difficult to study experimentally and theoretically \citep{ChamelHaensel2008}. Most observational signals sensitive 
to the crust do not isolate its contribution. Their interpretation depends on its coupling to the core,
 superfluid components, magnetic field, and thermal structure
\citep{ChamelHaensel2008,Andersson2012Glitches,Chamel2013Entrainment,PageReddy2013}.

Over the past decade, nuclear theory, laboratory experiments, and
multimessenger observations have substantially improved our knowledge of the NS
equation of state (EOS)
\citep{Essick2020Nonparametric,Annala2020Quark,Raaijmakers2021EOS,
Pang2021Multimessenger,Huth2022Nature,Tsang2024DataCentric}.  Nuclear masses
and neutron skins probe neutron-rich matter below saturation density,
microscopic many-body calculations constrain matter up to about saturation density,
and heavy-ion collisions extend the laboratory reach to higher densities.
Mass measurements of massive pulsars, mass--radius measurements from the
Neutron Star Interior Composition Explorer (NICER), and tidal information
from binary mergers provide complementary constraints on cold matter in NS
cores.  These inputs have been combined in a wide range of EOS studies based 
on different descriptions of dense matter and assumptions about its composition
and phase structure \citep{Fortin2016,Annala2020Quark, Tsang2024DataCentric, Huth2022Nature}.
Many such studies focus on the core and
attach a standard crust EOS to each sampled core EOS.  This approximation has
little effect on the maximum mass and only a modest effect on other global
properties.  It can nevertheless change stellar radii at the few-percent
level, especially at low mass, and produce much larger errors in the crust
thickness \citep{Fortin2016,Ferreira2020Matching, ParmarBombaci2026}.
A consistent crust--core description is particularly important for estimating
the radii of low-mass neutron stars and the relation between their baryonic
and gravitational masses \citep{ParmarScordinoBombaci2026LowMass}.

A fixed crust may therefore be adequate for many studies of the core EOS, but
it cannot be used to determine what constrains the crust itself.  The
crust--core transition is set below saturation density by the competition
between the bulk energy of neutron-rich matter and the surface and Coulomb
energies of nuclear clusters.  To leading order, the transition pressure sets
the crust mass and crustal moment of inertia, while the difference in baryon chemical potential
between the stellar surface and the crust--core boundary determines the
crust thickness
\citep{Carreau2019Transition,Carreau2019Moment,Newton2022Skins}.  At the same
time, these stellar crust properties depend on the mass, radius, and therefore
the EOS of the whole star.  A crust calculation must consequently be embedded
in a complete crust--core EOS, on which nuclear and astrophysical constraints
act together.  Moreover, the part of the EOS tested by a global observable
depends on the stellar mass.  \citet{Weih2019OptimalMass}, for example, showed
that measurements at different masses do not constrain the same density
region, and that intermediate-mass stars can provide greater leverage on the
lower-density EOS than the most massive stars. 

Several Bayesian studies have treated the crust and core consistently or have focused specifically on crust properties
\citep{Carreau2019Transition,Carreau2019Moment,Balliet2021,DinhThi2021,
Newton2022Skins,Scurto2024,Burrello2025,ParmarBombaci2026}.
These studies show that microscopic neutron-matter calculations and the poorly known surface energy of very neutron-rich clusters strongly affect the crust--core transition
\citep{Carreau2019Transition,Carreau2019Moment}.
Nuclear masses, neutron-skin measurements, and finite-size terms further constrain the transition and the pasta layers
\citep{Balliet2021,DinhThi2021,Newton2022Skins}.
However, the relative contributions of individual nuclear and astrophysical
constraints, and the extent to which future stellar measurements can replace
low-density nuclear information, remain less systematically explored.

Here, we address these questions using Bayesian inference within a unified
crust--core model.  We compare combinations of present constraints,
variations in the nuclear model, and forecasts across stellar masses and
measurement precisions.  We trace how the inferred crust--core transition, crust thickness, 
crust mass, and crustal moment of inertia respond when individual nuclear and astrophysical
constraints are removed.  We focus in particular on chiral effective field theory ($\chi\mathrm{EFT}$) 
because previous Bayesian studies have found microscopic neutron-matter calculations to 
be among the strongest constraints on the crust--core transition and related crust
properties \citep{Carreau2019Transition,Carreau2019Moment,Newton2022Skins,Burrello2025}. 
We then examine whether precise measurements of radius, tidal deformability, and total moment 
of inertia can provide comparable information across different stellar masses. Finally, we test 
whether precise global measurements can distinguish different crusts.
We identify two unified EOSs with almost identical radii, tidal
deformabilities, and maximum masses, but with well-separated crust
properties.  We then generate mock radius, tidal-deformability, and
moment-of-inertia measurements from each EOS, using the same uncertainties,
and infer the crust in the full model.  These ``crust doppelg\"angers'' extend
the EOS-degeneracy studies of
\citet{RaithelMost2023PRL,RaithelMost2023PRD} to the crust itself.

\section{Unified crust and core model}
\label{sec:model}

We describe the inner crust with the compressible liquid-drop model (CLDM) \citep{Carreau2019Transition,Carreau2019Moment}.
  In the CLDM, matter is described within a spherical Wigner--Seitz cell containing a neutron-rich nuclear cluster, electrons, 
  and unbound neutrons. At each mean density, the equilibrium composition is found by minimizing the total energy, 
  including bulk, surface, curvature, and Coulomb contributions. The competition between these contributions determines whether clustered 
  or uniform matter is favoured and thus sets the crust--core transition. For the outer crust, we use experimental nuclear masses where available.
   In the deeper layers, where increasingly neutron-rich nuclei appear and measured masses are unavailable, we use the HFB-24 mass model \citep{Goriely2013HFB24}. 
   We keep this outer-crust EOS fixed for all sampled EOSs.

The bulk energies of the cluster, neutron gas, and uniform core are supplied
by the empirical metamodel of
\citet{Margueron2018I,Margueron2018II}.  The interaction contribution to the
energy per baryon is expanded to fourth order around saturation density.
A low-density correction ensures that the energy per baryon vanishes at
zero density.  The parameter vector $\bm\theta$
contains thirteen empirical parameters,
\begin{equation*}
 \begin{split}
 \bm{\theta}=\{&\nsat,\Esat,\Ksat,\Qsat,\Zsat,
 \Jsym,\Lsym,\Ksym,\\
 &\Qsym,\Zsym,m^{\ast}/m,\Delta m^{\ast},b\}.
 \end{split}
\end{equation*}

Here, $\nsat$ is the saturation density of symmetric nuclear matter, and $\Esat$ 
is its energy per baryon at saturation. The coefficients $\Ksat$, $\Qsat$, 
and $\Zsat$ are the incompressibility, skewness, and kurtosis parameters, 
respectively, and characterize the second-, third-, and fourth-order density
 dependence of the energy per baryon around $\nsat$. In the isovector sector,
  $\Jsym$ is the symmetry energy at saturation, while $\Lsym$, $\Ksym$, $\Qsym$, 
  and $\Zsym$ are its slope, curvature, skewness, and kurtosis parameters, respectively, 
  evaluated at $\nsat$. The ratio $m^{\ast}/m$ is the nucleon effective mass in symmetric matter at
saturation divided by the bare nucleon mass $m$.  The dimensionless splitting
$\Delta m^{\ast}=(m_n^{\ast}-m_p^{\ast})/m$ is evaluated in pure-neutron
matter at saturation, where $m_n^{\ast}$ and $m_p^{\ast}$ are the neutron
and proton effective masses.  These two parameters control the
momentum-dependent kinetic energy, while the dimensionless parameter $b$
controls the low-density correction.
 In benchmarks against existing nuclear models, \citet{Margueron2018I}
showed that the fourth-order metamodel can reproduce a broad range of
smooth nucleonic EOSs when its high-order coefficients are adjusted
to their density dependence.
Bayesian calculations with relativistic mean-field models and nonrelativistic
metamodels have yielded similar crust-property distributions when low-density
nuclear constraints are included \citep{Scurto2024,ParmarBombaci2026}.  We use the metamodel
here because its computational efficiency makes the repeated inference
across constraint combinations, model variants, and forecast scenarios
practical.
The independent uniform priors adopted for these parameters are listed in
Table~\ref{tab:priors} of Appendix~\ref{app:present-likelihood}.

For each sampled EOS, the surface and curvature coefficients are
fitted anew to the masses from the 2020 Atomic Mass Evaluation (AME2020)
\citep{Wang2021AME2020}.  This keeps the surface and bulk contributions consistent for each EOS.  We adopt the
surface-tension parametrization introduced by \citet{Ravenhall1983Surface}.
It is given in Eq.~\eqref{eq:surface} of Appendix~\ref{app:model}, where the
exponent $p$ controls the extrapolation to small proton fractions.  We adopt
$p=3$, a commonly used reference choice in liquid-drop crust models
\citep{Newton2013CLDM,Carreau2019Transition,Grams2022CLDMETF}.  This choice reproduces dynamical-spinodal
transition densities reasonably well for several nuclear interactions
\citep{Carreau2019Transition}.  Comparisons with extended Thomas--Fermi
calculations also favour values close to three, although the preferred value
depends on the nuclear interaction and treatment of the surface
\citep{Grams2022CLDMETF}.  We therefore repeat the calculation with $p=2.5$
and $3.5$ to assess this model dependence. 

The nonuniform crust and uniform core use the same nuclear interaction.  The
crust--core transition occurs where uniform matter becomes energetically
favoured over the optimized nuclear configuration.  We retain only EOSs that
are stable, thermodynamically consistent, and causal.  A one-sided mass
constraint suppresses EOSs whose maximum mass $\Mmax$ falls below $2\,\Msun$,
where $\Msun$ is the solar mass (Appendix~\ref{app:inference}).  We collect the five
crust quantities studied here in the vector $\bm y_{\rm crust}$,
\begin{equation}
 \bm{y}_{\rm crust}=\{n_t,P_t,\Delta R_{\rm crust},
 M_{\rm crust},I_{\rm crust}/I\},
 \label{eq:crusttargets}
\end{equation}
where $n_t$ and $P_t$ are the baryon density and pressure at the crust--core
transition.  The crust thickness is $\Delta R_{\rm crust}=R-R_{\rm core}$,
and its gravitational mass is $M_{\rm crust}=M-M_{\rm core}$, where $M$ and
$R$ are the stellar gravitational mass and radius, and $M_{\rm core}$ and
$R_{\rm core}$ are their values at the transition.  The ratio
$I_{\rm crust}/I$ is the crustal contribution $I_{\rm crust}$ to the total
stellar moment of inertia $I$.  These three stellar crust quantities depend
on $M$.  We evaluate them at $1.4\,\Msun$ unless stated otherwise.
Mass arguments such as $(1.4)$ denote stellar masses in units of $\Msun$.  These quantities describe the crust--core boundary and the crust contributions
to the stellar radius, mass, and moment of inertia.
The metamodel expansion and CLDM energy contributions are summarized in
Appendix~\ref{app:model}.  The priors and likelihoods are listed in
Appendix~\ref{app:inference}.

\section{Present constraints and forecast construction}
\label{sec:bayes}

\subsection{Present nuclear and astrophysical constraints}

The present constraints comprise nuclear-matter calculations, laboratory
measurements, and neutron-star observations, which probe different density
ranges and EOS properties.  The combined likelihood is written as
\begin{equation}
 \mathcal{L}_{\rm present}=\mathcal{L}_{\rm phys}\mathcal{L}_{\rm sat}
 \mathcal{L}_{\rm lab}\mathcal{L}_{\chi{\rm EFT}}\mathcal{L}_{\Mmax}
 \mathcal{L}_{\rm NICER}\mathcal{L}_{\rm GW170817}.
 \label{eq:factorization}
\end{equation}
The physical filter $\mathcal{L}_{\rm phys}$ enforces thermodynamic
consistency, stability, positive symmetry energy, and causality throughout
the unified EOS.  The saturation term $\mathcal{L}_{\rm sat}$ constrains the saturation
density, energy per baryon, and incompressibility of symmetric nuclear
matter \citep{Buchinger1994Radii,Chabanat1997SLy,Khan2012Incompressibility,
RocaMaza2015Covariance,Margueron2018I}.  Additional laboratory
constraints at finite density enter through
$\mathcal{L}_{\rm lab}$ and constrain the symmetry energy, its density
dependence, and the pressure of symmetric matter.  These inputs are drawn
from the experimental compilation of \citet{Tsang2024DataCentric}.  The
individual measurements and their original sources are listed in
Table~\ref{tab:nuclear-constraints}.

The low-density EOS is constrained by \chEFT{} calculations of symmetric
and pure-neutron matter between $0.02$ and $0.20\,\mathrm{fm}^{-3}$
\citep{Drischler2016,Hebeler2013}.  This interval covers densities relevant to
the inner crust and the transition to uniform matter, making \chEFT{} the
principal microscopic constraint in this regime.  The factor
$\mathcal{L}_{\chi{\rm EFT}}$ assigns equal weight to EOSs whose energies lie
within the \chEFT{} bands at the evaluated densities in this interval and rejects
those outside.  In the reference calculation, the band boundaries are
widened by 5\%.  Sensitivity to this band-widening prescription is assessed
using the original bands and widenings of
10\% and 20\%.  A separate assessment of regulator-cutoff dependence and
order-by-order \chEFT{} truncation uncertainties is beyond the scope of this
work (see \citealt{Goettling2026} for a quantified truncation-error treatment).
The percentage widenings mimic a broader theoretical uncertainty band and
test its effect on the inferred crust properties, but do not provide a
statistical estimate of these uncertainties.
Calculations without the \chEFT{} constraint assess the
independent constraining power of laboratory and stellar data.

The astrophysical factors constrain the global stellar structure.
The term $\mathcal{L}_{\Mmax}$ constrains the maximum supported
gravitational mass $\Mmax$ through a smooth penalty below
$2\,\Msun$, motivated by the observed massive neutron stars \citep{Antoniadis2013,Cromartie2020}.
The NICER term includes the mass--radius constraints for
PSRs J0030+0451 \citep{Riley2019J0030},
J0437--4715 \citep{Choudhury2024J0437},
J0614--3329 \citep{Mauviard2025J0614}, and
J0740+6620 \citep{Riley2021J0740}.
For each source, the measured mass--radius distribution is evaluated along
the predicted $R(M)$ curve, retaining the uncertainty in stellar mass.
The GW170817 term uses the dimensionless binary tidal deformability
$\widetilde\Lambda$ inferred by \citet{Abbott2018GW170817EOS}, with the chirp
mass fixed and the mass ratio marginalized.  This quantity combines the
individual dimensionless tidal deformabilities $\Lambda(M)$ according to
Eq.~\eqref{eq:lambdatilde}.  The adopted constraints, their
sources, and the likelihood definitions are given in
Appendix~\ref{app:present-likelihood}.

\begin{figure*}[!t]
\centering
\includegraphics[width=\textwidth]{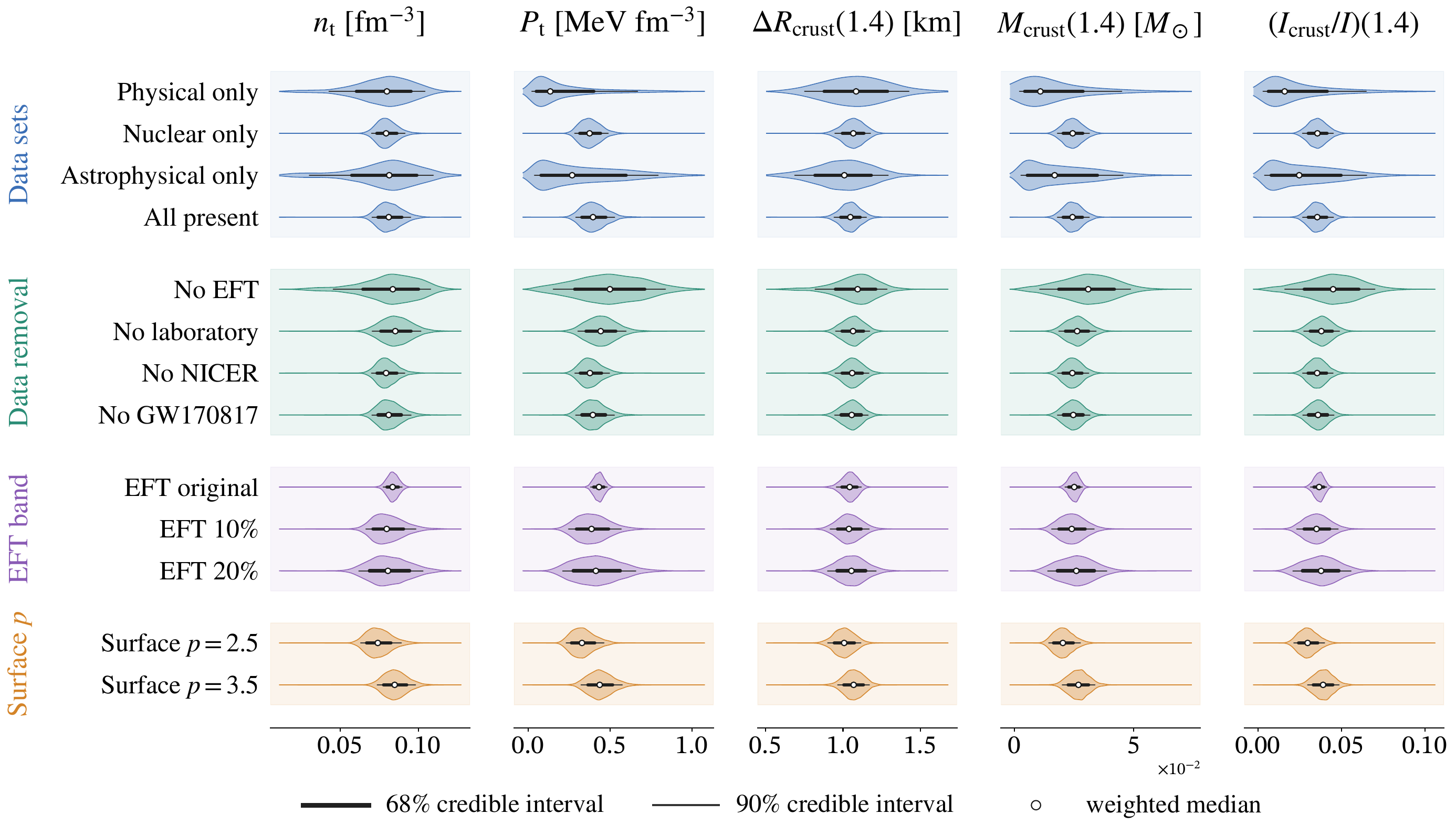}
\caption{Present constraints on the five crust quantities.  The first group
of rows compares the physical baseline with the nuclear-only,
astrophysical-only, and full present-data calculations.  The remaining groups
show the effects of removing individual constraints, varying the \chEFT{}
envelope, and changing the surface exponent.  The shaded shapes show
marginalized probability densities.  Thin and thick horizontal segments mark
the central 90\% and 68\% credible intervals, respectively.  Open circles
mark the medians.}
\label{fig:present-posteriors}
\end{figure*}

\subsection{Forecasts of future measurements}
\label{sec:forecast-construction}

Measurements of stellar radius, tidal deformability, and moment of inertia
probe the EOS through the global structure of the star.  We assess how much
improved measurements of these observables can reduce crust uncertainties
beyond the present constraints, and whether this information depends on the
low-density \chEFT{} input. 

We construct mock measurements using a single reference EOS, $T_{\rm REF}$,
selected from the full present-data posterior near the nuclear-parameter
medians.  For each observable and stellar mass, we take this EOS prediction
as the central measured value and assign the uncertainty specified below.
For example, a 2\% radius forecast is centred on the predicted radius and
has a standard deviation equal to 2\% of that value.  We then combine the
mock measurements with the present constraints to infer the crust properties.

The central values are not randomly displaced by measurement noise.  These
zero-noise, or Asimov, data \citep{Cowan2011Asimov} allow comparisons of
measurement precision and combinations of observables for the same assumed
underlying EOS.  The forecast gains are conditional on this reference EOS and are not
averaged over all EOSs allowed by present data.  Systematic errors from instrument calibration or from approximations in
the gravitational-wave models used to infer tidal deformability are not
included.  Appendix~\ref{app:forecast-likelihood} gives further details.

The radius forecasts span $1.2$--$1.8\,\Msun$ in steps of $0.2\,\Msun$,
with 2\% uncertainties in both mass and radius, following the precision
benchmark used by \citet{Huang2025} for future X-ray measurements.
Adopting the same fractional precision at each mass isolates the mass
dependence of the constraints.  We label a binary forecast by
$\widetilde\Lambda_{m_1+m_2}$, with the component masses in solar units.
Tidal forecasts assume a 10\% uncertainty for equal-mass binaries with
component masses of $1.2$, $1.4$, and $1.6\,\Msun$.  We also test 5\% and
20\% uncertainties for the $1.4+1.4\,\Msun$ binary.  These precisions are
motivated by prospects for next-generation gravitational-wave detectors
\citep{Iacovelli2023ET}, but serve as benchmarks rather than predictions
for a specific event.  For the moment of inertia of PSR J0737--3039A,
with mass $1.338\,\Msun$, we adopt the 10\% precision considered by
\citet{Lim2019Inertia} and a more optimistic 5\% case.  Combined forecasts test the complementarity of the
three observables.  The ``realistic combined'' case uses a 2\% mass--radius
measurement at $1.8\,\Msun$, a 10\% tidal measurement for a
$1.4+1.4\,\Msun$ binary, and a 10\% measurement of $I(1.338\,\Msun)$.
The ``multi-star combined'' case replaces the single mass--radius measurement
with five at $1.2$, $1.4$, $1.6$, $1.8$, and $2.0\,\Msun$, each at the same
2\% precision.  We also consider these five mass--radius measurements alone.

Selected forecasts are repeated without the \chEFT{} constraint, keeping
the reference EOS and forecast measurements fixed.  This comparison isolates
the information supplied by future observations independently of the
low-density \chEFT{} input.

\section{Statistical analysis}
\label{sec:design}

For each crust quantity $X$, we denote its weighted posterior $a$th
percentile by $q_a(X)$ and use it to define the central widths $W_{68}=q_{84}-q_{16}$ and
$W_{90}=q_{95}-q_5$.  Present-data comparisons use
$W_{68}^{\rm case}/W_{68}^{\rm ref}$, with values above unity indicating
broader intervals.  Forecasts use the inverse ratio,
\begin{equation}
 G_X=\frac{W_{68}^{\rm ref}(X)}{W_{68}^{\rm case}(X)}.
 \label{eq:contraction}
\end{equation}
Values of $G_X$ above unity indicate narrower credible intervals, while
values below unity indicate broader intervals relative to the reference case.  The geometric mean,
$G_{\rm geom}=(\prod_XG_X)^{1/5}$, summarizes the width changes across the five crust quantities.
Changes in posterior location and shape are additionally quantified by the
marginal Kullback--Leibler divergence $D_{\rm KL}^{X}$ \citep{KullbackLeibler1951},
\begin{equation}
 D_{\rm KL}^{X}({\rm case}\Vert{\rm ref})=
 \int p_{\rm case}(X)\log_2\!\left[
 \frac{p_{\rm case}(X)}{p_{\rm ref}(X)}\right]\dd X.
 \label{eq:dkl}
\end{equation}
Here $p_{\rm case}(X)$ and $p_{\rm ref}(X)$ are the normalized marginal
posterior densities for the case and reference calculations.  The base-two
logarithm expresses $D_{\rm KL}^{X}$ in bits.  It is zero for identical
distributions and positive otherwise.  It is not symmetric under
exchange of the two distributions and is not a detection significance.

\begin{figure*}[!t]
\centering
\includegraphics[width=\textwidth]{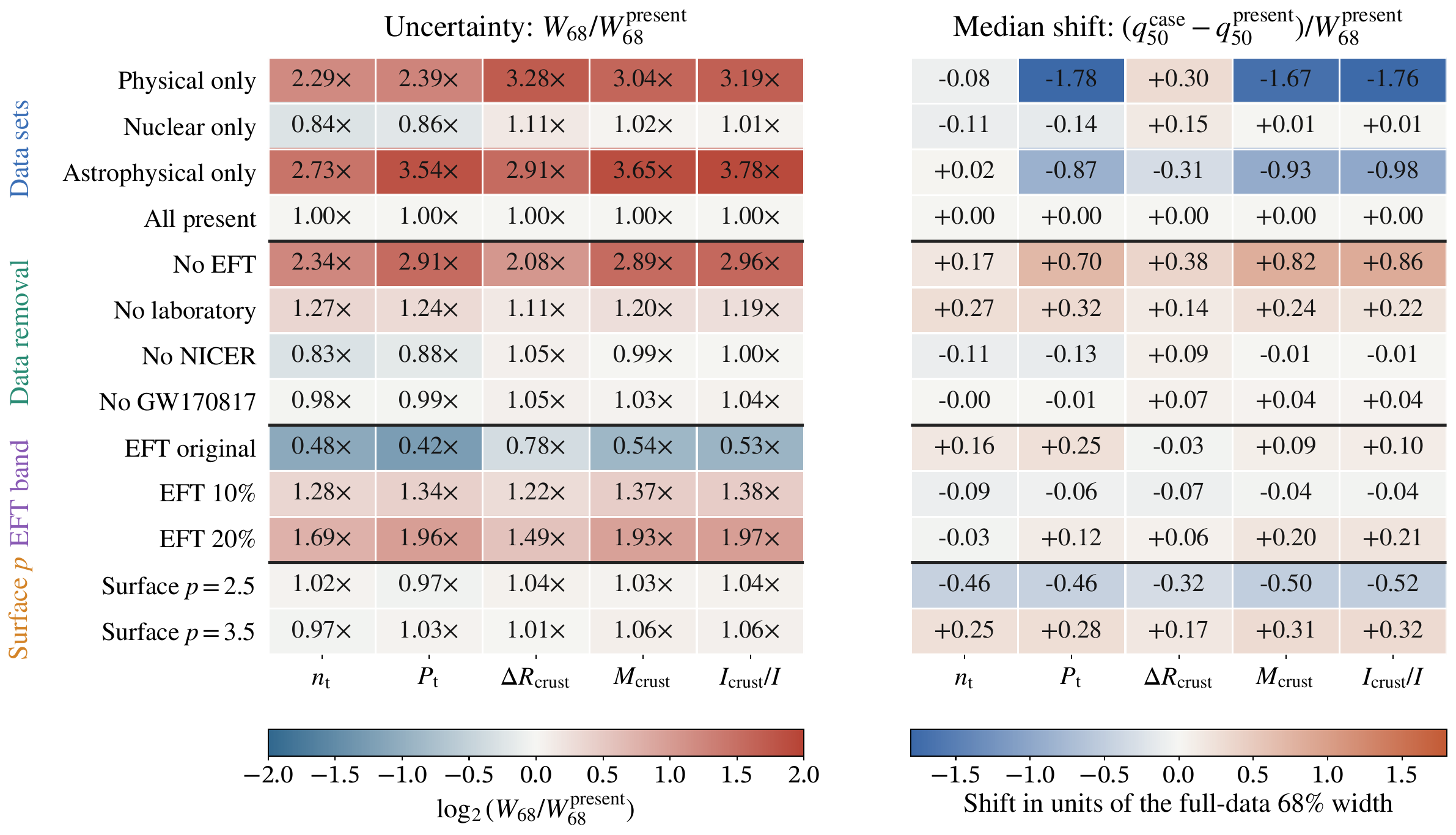}
\caption{Sensitivity of inferred crust properties to present constraints and
nuclear modelling.  All cases are compared with the full present-data
calculation (``All present''), which uses a 5\% \chEFT{}-band widening and $p=3$.
\emph{Left:} ratio of the central 68\% credible-interval width to its
reference value.  Numbers give the ratios, while colours show their
base-two logarithms.  Red indicates broader intervals and blue narrower
intervals.  A value of unity indicates no change in width.
\emph{Right:} change in the median divided by the reference 68\% width.
Positive (red) and negative (blue) values indicate shifts to higher and
lower values, respectively.  The first group of rows compares the physical, nuclear,
and astrophysical constraints.  The remaining groups show the effects of
omitting individual inputs, varying the \chEFT{}-band widening, and changing
the surface exponent $p$.
Crust thickness, mass, and fractional moment of inertia are evaluated at
$1.4\,\Msun$.}
\label{fig:present-effects}
\end{figure*}

For forecasts without \chEFT{}, we also define
\begin{equation}
\begin{aligned}
 G_X^{\rm obs}&=\frac{W_{68}^{\rm no\ \chi EFT}(X)}
 {W_{68}^{\rm no\ \chi EFT+future}(X)},\\
 R_X^{\rm cur}&=\frac{W_{68}^{\rm no\ \chi EFT+future}(X)}
 {W_{68}^{\rm present}(X)},\qquad
 R_X^{\rm match}=\frac{W_{68}^{\rm no\ \chi EFT+future}(X)}
 {W_{68}^{\rm present+future}(X)}.
\end{aligned}
 \label{eq:noeft-gains}
\end{equation}
Here $G_X^{\rm obs}$ measures narrowing relative to the present-data
posterior without \chEFT{}.  The ratios $R_X^{\rm cur}$ and $R_X^{\rm match}$
compare the resulting width with the full present-data posterior and the
corresponding \chEFT{}-conditioned forecast, respectively.  Unity therefore
indicates recovery of the present precision for $R_X^{\rm cur}$, or equal
forecast widths with and without \chEFT{} for $R_X^{\rm match}$.

Finally, we select pairs from the full present-data posterior with similar
$R(M)$, $\Lambda(M)$, and maximum mass but different crust properties.
Moment of inertia is excluded from the selection to retain it as an
independent test.  Forecast recoveries centred on each member test whether
future observations can distinguish these crusts.  This \emph{crust-doppelg\"anger} test is motivated by the analysis of
\citet{RaithelMost2023PRL,RaithelMost2023PRD}.  The selection and recovery
criteria are defined in Sect.~\ref{sec:doppel-results}, with supporting
details in Appendix~\ref{app:doppel}.

These tests measure distinguishability at the assumed precision in observable
space.  Detector-specific claims would additionally require waveform and
noise modelling, source populations, and selection effects
\citep{Hotokezaka2016Measurability,Pacilio2022Ranking,Iacovelli2023ET}.

\section{Results and discussion}
\label{sec:results}

\subsection{Present constraints on crust properties}

We compare thirteen configurations of the constraints and nuclear model to assess how
nuclear and stellar data constrain the crust.  The physical baseline includes
the admissibility conditions and the lower bound on maximum mass.  The
nuclear-only case adds saturation, laboratory, and \chEFT{} constraints,
while the astrophysical-only case adds NICER and GW170817.
Figure~\ref{fig:present-posteriors}
shows the corresponding distributions of the five crust quantities.  With
all present constraints included, their medians and central 68\% credible
intervals are
\begin{equation}
\begin{aligned}
n_t &= 0.0811^{+0.0084}_{-0.0069}\ {\rm fm}^{-3},\\
P_t &= 0.396^{+0.078}_{-0.068}\ {\rm MeV\,fm}^{-3},\\
\Delta R_{\rm crust}(1.4) &= 1.048^{+0.063}_{-0.062}\ {\rm km},\\
M_{\rm crust}(1.4) &= 0.0243^{+0.0041}_{-0.0039}\ \Msun,\\
(I_{\rm crust}/I)(1.4) &= 0.0355^{+0.0057}_{-0.0054}.
\end{aligned}
\label{eq:present-headline}
\end{equation}

Table~\ref{tab:appendix-fiducial-posteriors} also lists the nuclear-parameter
posteriors and the central 90\% intervals.  These intervals describe the
uncertainty within the adopted nuclear model and constraints.  Nuclear
constraints dominate the present inference: relative to the
physical baseline, laboratory data together with \chEFT{} give width ratios of
0.318--0.365.  Including stellar observations gives ratios of 0.305--0.436.
Stellar data alone narrow only the crust thickness, to 0.888 times its
baseline width.  The other four central intervals become broader as the
posterior shapes change.  Without nuclear information, current stellar
observations therefore provide limited constraints on the transition, crust
mass, and crustal moment of inertia.

Figure~\ref{fig:present-effects} summarizes the changes in widths and
medians.  Without \chEFT{}, the five widths increase by factors of 2.08--2.96,
because a wider range of neutron-rich matter properties is then allowed
near the crust--core transition.  Removing both saturation and finite-density laboratory constraints increases them by factors of
1.11--1.27, whereas omitting NICER or GW170817 gives most width ratios
between 0.95 and 1.05.  Some transition intervals become narrower without NICER because
removing NICER changes the shape of the posterior.  This change does not by itself indicate
a stronger constraint. The inferred precision is also sensitive to the adopted \chEFT{} band width.
Relative to the reference 5\% widening, the
original band gives width ratios of 0.42--0.78.  Widenings of 10\% and 20\%
give ratios of 1.22--1.38 and 1.49--1.97, respectively, with smaller
changes in the medians.

Changing the neutron-rich surface energy shifts the inferred crust properties
while leaving their credible widths nearly unchanged.  Relative to $p=3$, setting $p=2.5$ lowers the medians by
0.32--0.52 reference widths, while $p=3.5$ raises them by 0.17--0.32 widths.
The separation between the two choices reaches 0.49--0.84 reference widths,
although the central 68\% width ratios remain close to unity
(Table~\ref{tab:appendix-present-effects}). This sensitivity to the neutron-rich surface energy is consistent with
earlier crust calculations.  \citet{Newton2013CLDM} showed that the surface
energy at small proton fractions strongly affects the inner-crust composition
and transition density.  \citet{Carreau2019Transition} subsequently found
that increasing $p$ raises both the transition density and pressure, as in
our results, and that uncertainty in $p$ broadens their inferred ranges.
The importance of treating the surface and bulk contributions consistently
was also demonstrated by \citet{DinhThi2021}. The shifts between surface prescriptions are a substantial fraction of the
present uncertainty, which is not captured by the intervals at fixed $p$.
The three choices of $p$ test this sensitivity without providing a full
uncertainty estimate for the surface functional.  Joint inference of bulk and surface parameters offers
a way to include this additional uncertainty \citep{Klausner2026}.

\begin{figure*}[t]
\centering
\includegraphics[width=\textwidth]{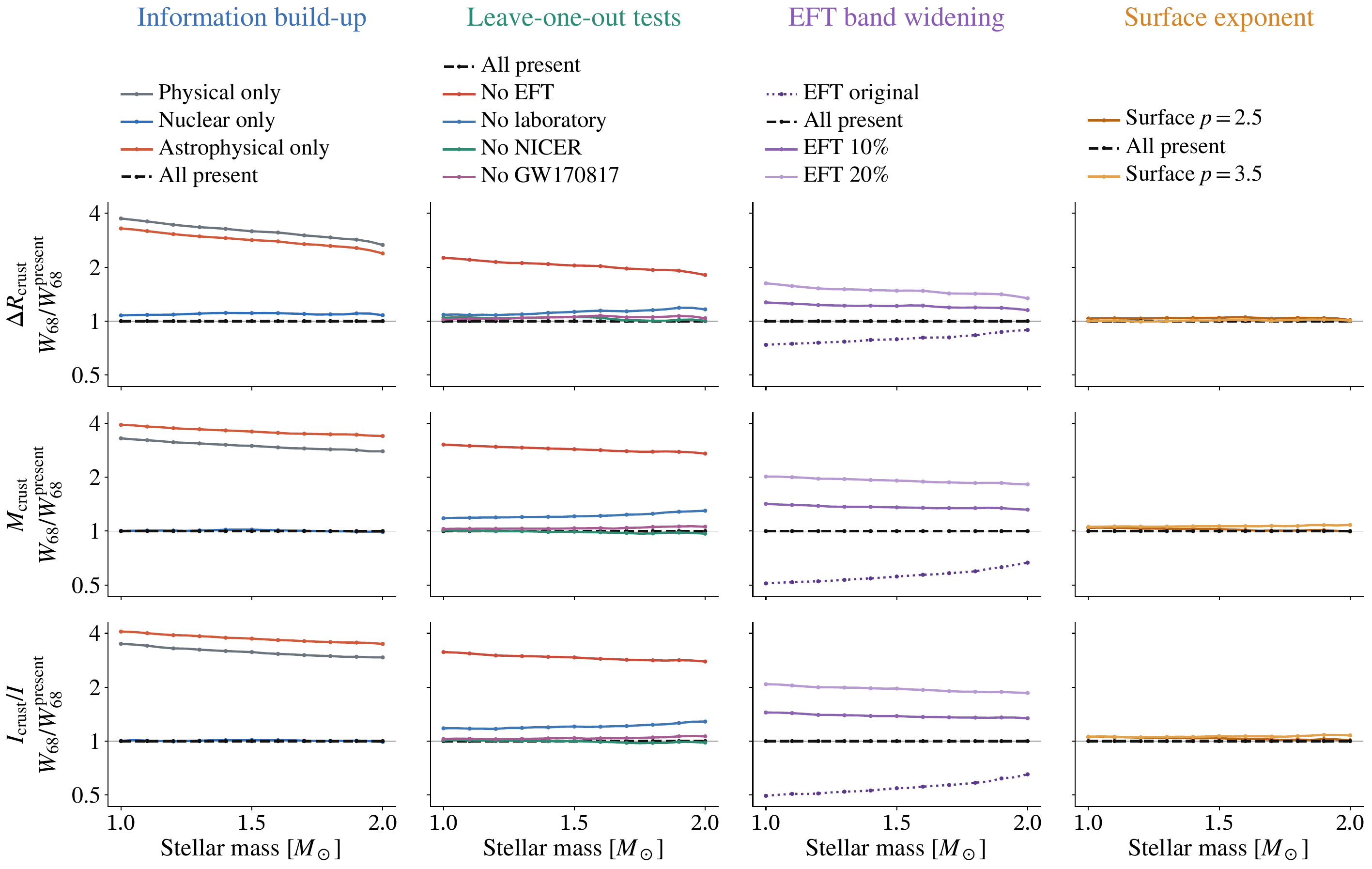}
\caption{Mass dependence of the central 68\% widths for crust
thickness (top), crust mass (middle), and fractional crustal moment of inertia
(bottom).  Each curve is divided by the full present-data width at the same
stellar mass.  The columns show the information build-up, leave-one-out tests,
\chEFT{}-envelope changes, and surface-energy models over
$1.0$--$2.0\,\Msun$.}
\label{fig:present-mass-robustness}
\end{figure*}

The same trends are seen across stellar masses from $1.0$ to $2.0\,\Msun$,
as shown in Figure~\ref{fig:present-mass-robustness}.  The corresponding crust
properties are listed in Table~\ref{tab:appendix-mass-dependent}.
Over this mass range, removing \chEFT{} increases the widths by factors of
1.81--3.14, while removing laboratory constraints gives factors of
1.08--1.30.  The width ratios remain close to unity when NICER or GW170817
is omitted (about 0.93--1.07) or the surface model is changed
(about 0.92--1.08).

The inferred transition agrees with earlier unified calculations.
\citet{Carreau2019Transition} obtained
$n_t=0.072\pm0.011\,{\rm fm}^{-3}$ and
$P_t=0.339\pm0.115\,{\rm MeV\,fm}^{-3}$, while
\citet{Newton2022Skins} found $P_t=0.33\pm0.07\,{\rm MeV\,fm}^{-3}$ from
neutron skins and \chEFT{} neutron matter.  \citet{Burrello2025} obtained $n_t=0.074\pm0.014\,{\rm fm}^{-3}$,
while \citet{Goettling2026} found $0.062$--$0.088\,{\rm fm}^{-3}$.
These studies use different surface models and treatments of theoretical
uncertainty.  They nevertheless favour a similar transition-density range,
while their pressure estimates vary more. The leading contribution from low-density theory and the additional
constraint from laboratory data are also consistent with previous work
\citep{Carreau2019Transition,Newton2022Skins,Burrello2025}.  Within the adopted unified model, the inferred crust properties are mainly
constrained by nuclear information and remain only weakly sensitive to the
inclusion of current NICER and GW170817 data.

\subsection{Constraints from future stellar measurements}

We first examine the improvement obtained by adding each forecast
measurement individually to the full present-data constraints, using the
scenarios defined in Sect.~\ref{sec:forecast-construction}.

\begin{figure*}[t]
\centering
\includegraphics[width=0.92\textwidth]{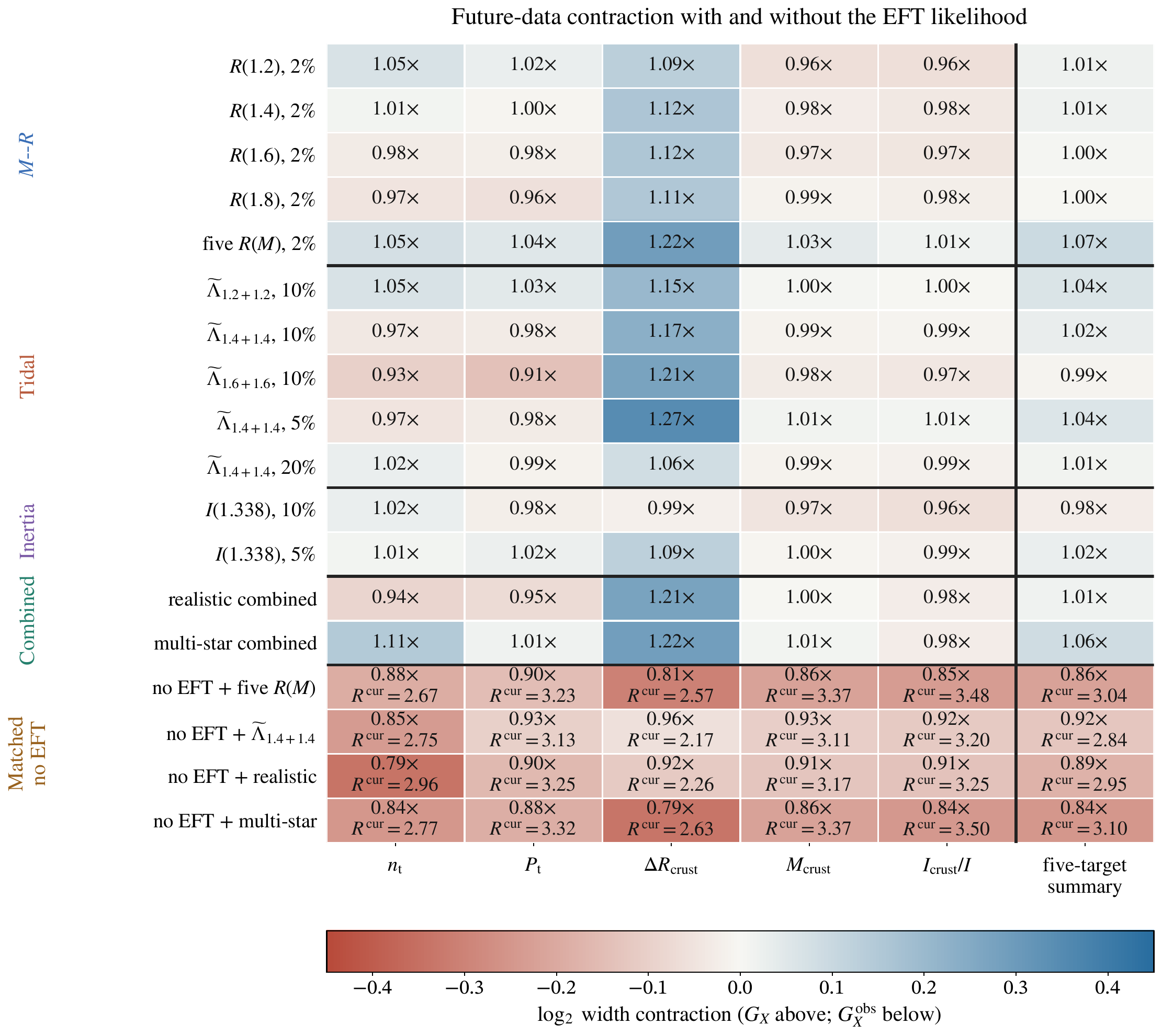}
\caption{Change in the central 68\% width after each idealized future
measurement.  The first fourteen rows show forecasts that retain
\chEFT{}, with $G_X=W_{68}^{\rm present}/W_{68}^{\rm present+future}$.  The last four rows show forecasts without
\chEFT{}.  In each of these cells, the upper number is
$G_X^{\rm obs}=W_{68}^{\rm no\ \chi EFT}/W_{68}^{\rm no\ \chi EFT+future}$ and the lower
number is
$R_X^{\rm cur}=W_{68}^{\rm no\ \chi EFT+future}/W_{68}^{\rm present}$.
Blue cells indicate narrower intervals and red cells broader intervals.  In
the last four rows, the colour represents $G_X^{\rm obs}$ only.  The final column
is the geometric mean over the five crust quantities.}
\label{fig:future-contraction}
\end{figure*}

Crust thickness shows the largest improvement among the five crust
quantities.  Its strongest constraint comes from the optimistic 5\%
measurement of the binary tidal deformability $\widetilde\Lambda$ for a
$1.4+1.4\,\Msun$ system, which reduces the central 68\% credible-interval
width by about 21\%.  Table~\ref{tab:appendix-future-effects} lists the width gains
and information gains for all forecast cases.  In Figure~\ref{fig:future-contraction}, individual 2\%
mass--radius measurements give thickness gains of 1.09--1.12.
The other width ratios remain within about 0.96--1.04, apart from a gain
of 1.05 in $n_t$ at $1.2\,\Msun$.  The geometric-mean gain is
1.00--1.01, with no clear advantage for any mass between $1.2$ and
$1.8\,\Msun$.

For the $1.4+1.4\,\Msun$ binary, relaxing the tidal uncertainty from 5\%
to 10\% and 20\% reduces the thickness gain from 1.27 to 1.17 and 1.06,
respectively.  The latter two correspond to width reductions of about 15\%
and 6\%.  The other four width ratios remain within about 0.97--1.03.
At fixed 10\% tidal uncertainty, the thickness gain increases from 1.15 for
the low-mass binary to 1.21 for the high-mass binary.  The latter broadens
both transition intervals (gains of 0.93 and 0.91), giving a
geometric-mean gain of 0.99, compared with 1.04 for the low-mass binary.
A 5\% moment-of-inertia measurement gives a thickness gain of 1.09.  The 10\% forecast leaves the widths nearly unchanged.

Across the fourteen forecasts, the median gain is 1.13 for thickness and lies
between 0.98 and 1.02 for every other quantity.  Of the fifteen gains that exceed 1.05 before rounding, thirteen are for
thickness and two for transition density.  All fifteen have central 95\% bootstrap intervals for $G_X$ above unity.  None concerns $P_t$, $M_{\rm crust}$, or $I_{\rm crust}/I$.
The limited
effect on the marginals is also reflected in the Kullback--Leibler divergence,
which compares the one-dimensional marginal posterior distributions.  Its largest value among
the 70 forecast comparisons is 0.16 bits.

Correlations between global observables and crust properties can help
interpret these results.  In the present-data posterior, we find that the
strongest Spearman rank-correlation coefficient $\rho$ between a global
observable and crust thickness reaches $|\rho|=0.69$, for
$I(1.6)$ and $\Delta R_{\rm crust}(1.4)$.  The corresponding maximum absolute
correlations with $n_t$, $P_t$, $M_{\rm crust}$, and $I_{\rm crust}/I$ are
0.40, 0.42, 0.37, and 0.34, respectively.
Among the nuclear parameters, $\Lsym$ correlates most strongly with $n_t$
and $P_t$ ($\rho=-0.75$ and $-0.59$), while $\Qsym$ has the strongest
correlation with the three stellar-crust quantities
($\rho=0.65$--0.67).  These correlations are consistent with the stronger forecast constraints on
crust thickness.

Table~\ref{tab:future-best-target} compares the best forecast for each
quantity with the changes caused by the nuclear model.  Only thickness reaches a gain
above 1.20.  The best gain for $n_t$ is 1.11, and the others do not exceed
1.04.  Relative to these forecast widths, removing \chEFT{} gives intervals
2.58--3.01 times as wide, while the difference between surface models is
0.62--0.84 widths.  The forecast improvement is therefore smaller than the
sensitivity to the low-density nuclear description.

\begin{table*}[t]
\caption{Forecast with the largest width contraction for each crust quantity.}
\label{tab:future-best-target}
\centering
\begin{tabular}{lccccc}
\toprule
Target & Best future case & $G_X$ & $D_{\rm KL}$ [bits]
& $\Delta_p/W_{68}^{\rm present+future}$
& $W_{68}^{\rm no\ \chi EFT}/W_{68}^{\rm present+future}$ \\
\midrule
$n_t$ & multi-star combined & 1.11 & 0.080 & 0.78 & 2.58 \\
$P_t$ & five $R(M)$, 2\% & 1.04 & 0.022 & 0.76 & 3.01 \\
$\Delta R_{\rm crust}(1.4)$ & $\widetilde\Lambda_{1.4+1.4}$, 5\% & 1.27 & 0.124 & 0.62 & 2.65 \\
$M_{\rm crust}(1.4)$ & five $R(M)$, 2\% & 1.03 & 0.064 & 0.83 & 2.98 \\
$(I_{\rm crust}/I)(1.4)$ & five $R(M)$, 2\% & 1.01 & 0.050 & 0.84 & 2.99 \\
\bottomrule
\end{tabular}
\tablefoot{The
gain is $G_X=W_{68}^{\rm present}/W_{68}^{\rm present+future}$, and $D_{\rm KL}$ is
measured relative to the full present-data posterior.  The last two columns
compare the forecast width with $\Delta_p=|q_{50}^{p=3.5}(X)-q_{50}^{p=2.5}(X)|$, the median
separation between the two surface models, and with the present width obtained without \chEFT{}.}
\end{table*}

\subsection{Combined measurements and residual model dependence}

To test whether combining measurements improves constraints on the other
crust properties, we consider five mass--radius measurements, a single
mass--radius measurement combined with tidal deformability and moment of
inertia (the ``realistic combined'' case), and all five mass--radius
measurements with the same tidal and inertia measurements (the ``multi-star
combined'' case).  The masses and uncertainties are specified in
Sect.~\ref{sec:forecast-construction}.  Each combination is added to the
full present-data constraints.

Comparisons at the same measurement precision show what each combination
adds (Table~\ref{tab:appendix-future-effects}).  The 10\% tidal-only
forecast for a $1.4+1.4\,\Msun$ binary gives a thickness gain of 1.17.
Adding the 2\% mass--radius measurement at $1.8\,\Msun$ and the 10\%
moment-of-inertia measurement raises this to 1.21.  The width reduction
relative to present data thus increases from about 15\% to 17\%, or an
additional narrowing of about 3\% relative to the tidal-only forecast.
Compared with the radius-only and inertia-only forecasts at these same
precisions, the combined thickness interval is about 8\% and 18\%
narrower, respectively.  For the other four crust quantities, the realistic
combination gives gains of 0.94--1.00 relative to present data and does not
produce a comparable improvement.

Combining five mass--radius measurements gives a thickness gain of 1.22,
compared with 1.09--1.12 for the individual radius forecasts considered here.
Adding the tidal and inertia measurements leaves this gain unchanged at
the reported precision.  The main additional improvement is in transition
density, whose gain rises from 1.05 to 1.11, corresponding to about 5\%
further narrowing.  The transition-pressure, crust-mass, and crustal-inertia
intervals instead become about 2--3\% wider.  Relative to the realistic
combination, the multi-star programme narrows the transition-density
interval by about 15\%, but the thickness interval by less than 1\%.
These comparisons show that combining observables does not substantially
extend the improvement beyond crust thickness, apart from a modest gain
in transition density.  They also show that the five-radius programme
already captures most of the thickness improvement in the combined forecast.
The stronger constraint on thickness is consistent with its close relation
to stellar radius and compactness, while the transition pressure, crust
mass, and crustal moment of inertia remain more dependent on sub-saturation
matter and surface physics.

\begin{figure*}[t]
\centering
\includegraphics[width=\textwidth]{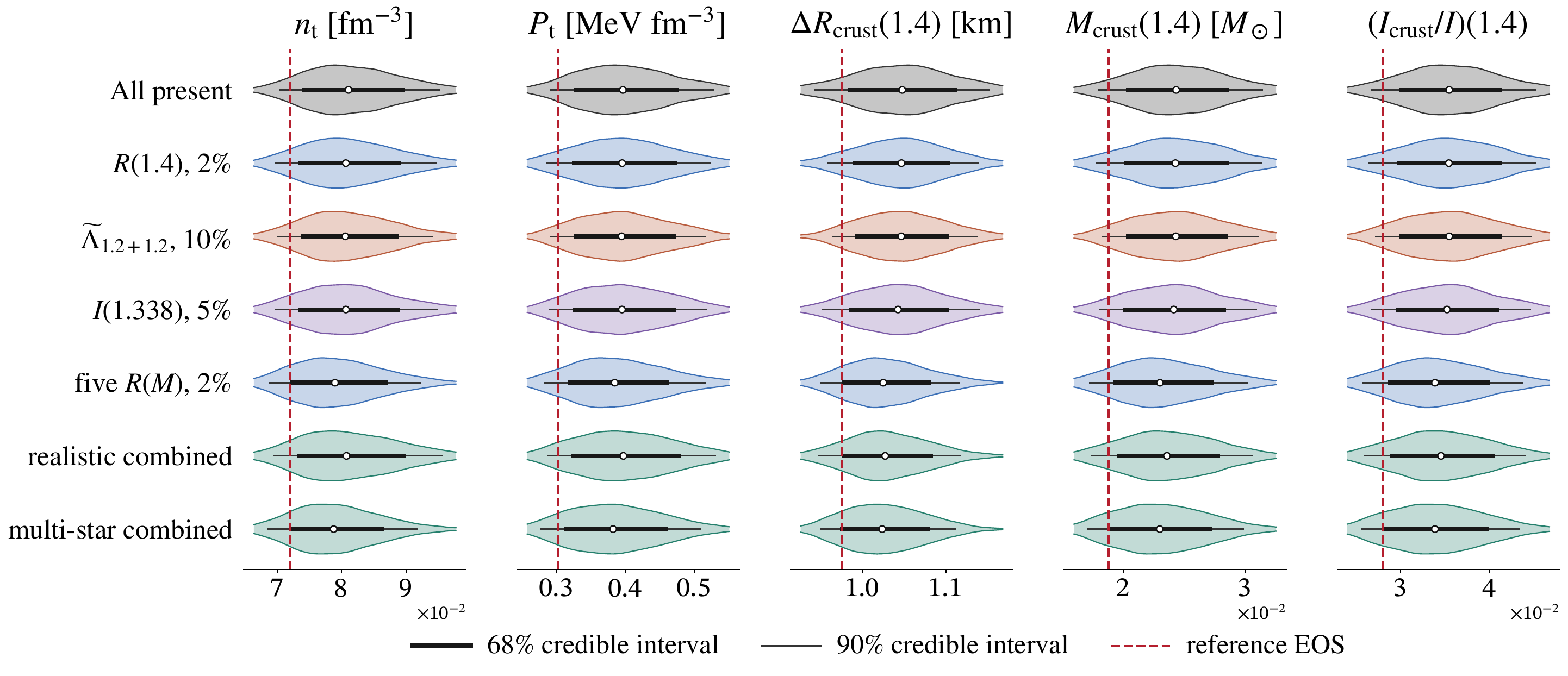}
\caption{Representative forecasts for the five crust quantities.  Selected
single radius, tidal, and moment-of-inertia measurements are shown
with the five-radius and combined programmes.  Thin and thick segments mark
the central 90\% and 68\% intervals, open circles mark the medians, and the
dashed red line marks the crust properties of the reference EOS used to generate all mock measurements.}
\label{fig:future-posteriors}
\end{figure*}

Figure~\ref{fig:future-posteriors} shows representative posteriors.
The reference crust lies within every central 90\% interval, but below
69 of the 70 central 68\% intervals.  The reference values lie at the 7th--12th percentiles of the present
posterior and at the 6th--17th percentiles after adding the forecast data.
Although selected near the nuclear-parameter medians, the reference EOS has
crust properties near the lower tail of the present posterior.  The forecast
measurements shift these distributions only slightly towards the reference values.

\subsection{Forecasts without the \chEFT{} constraint}

Removing \chEFT{} broadens the allowed low-density EOS range.  Four of the
forecasts are repeated with the same measurements but without \chEFT{} to test
whether stellar data can recover the lost precision.  The \chEFT{} constraint
extends to $0.20\,\mathrm{fm}^{-3}$, covering both the crust and part of the
uniform-matter regime.  Removing it therefore tests the combined contribution
of low-density nuclear theory, rather than isolating information about the
crust alone.  The comparison uses the width ratios defined in
Eq.~\eqref{eq:noeft-gains}.

None of the 20 central 68\% intervals narrows relative to the present-data
calculation without \chEFT{}: $G_X^{\rm obs}=0.79$--0.96
(Figure~\ref{fig:future-contraction}).  The full distributions
remain close to those obtained before adding the forecast measurements,
with $D_{\rm KL}\leq0.103$ bits.  Using central 90\% intervals also gives
no narrowing.  The wider intervals should not be interpreted as observations
reducing the available information.  They describe the response of this
interval measure to measurements centred on the chosen reference EOS.
We have not established whether the same behaviour persists for other
reference EOSs.

Without \chEFT{}, the forecast widths remain 2.17--3.50 times as wide as the
current widths obtained with \chEFT{} and 2.67--3.52 times as wide as the
corresponding forecasts retaining \chEFT{} (Table~\ref{tab:noeft-matched}).  Even a 5\%
tidal measurement leaves the thickness interval 2.17 times as wide as the
current result.  All central 90\% intervals contain the reference crust.

\begin{table}[t]
\caption{Crust uncertainty in forecasts without and with \chEFT{}.}
\label{tab:noeft-matched}
\centering
\small
\setlength{\tabcolsep}{3pt}
\begin{tabular}{@{}lccccc@{}}
\toprule
Forecast & $n_t$ & $P_t$ & $\Delta R_{\rm crust}$ & $M_{\rm crust}$ & $I_{\rm crust}/I$ \\
\midrule
Five $R(M)$, 2\% & 2.81 & 3.34 & 3.15 & 3.46 & 3.52 \\
$\widetilde\Lambda_{1.4+1.4}$, 5\% & 2.67 & 3.06 & 2.76 & 3.13 & 3.22 \\
Realistic combined & 2.79 & 3.09 & 2.72 & 3.17 & 3.19 \\
Multi-star combined & 3.07 & 3.35 & 3.21 & 3.39 & 3.44 \\
\bottomrule
\end{tabular}
\tablefoot{Entries give the ratio of the central 68\% width without
\chEFT{} to that with \chEFT{}, $R_X^{\rm match}$
(Eq.~\eqref{eq:noeft-gains}).  Both calculations include the same future
measurements and all other present constraints.
The stellar crust quantities are evaluated at $1.4\,\Msun$.}
\end{table}

\subsection{Distinguishing crust doppelg\"angers}
\label{sec:doppel-results}

The doppelg\"anger test asks whether EOSs with similar global stellar
properties can have distinguishable crusts.

\begin{figure*}[t]
\centering
\includegraphics[width=\textwidth]{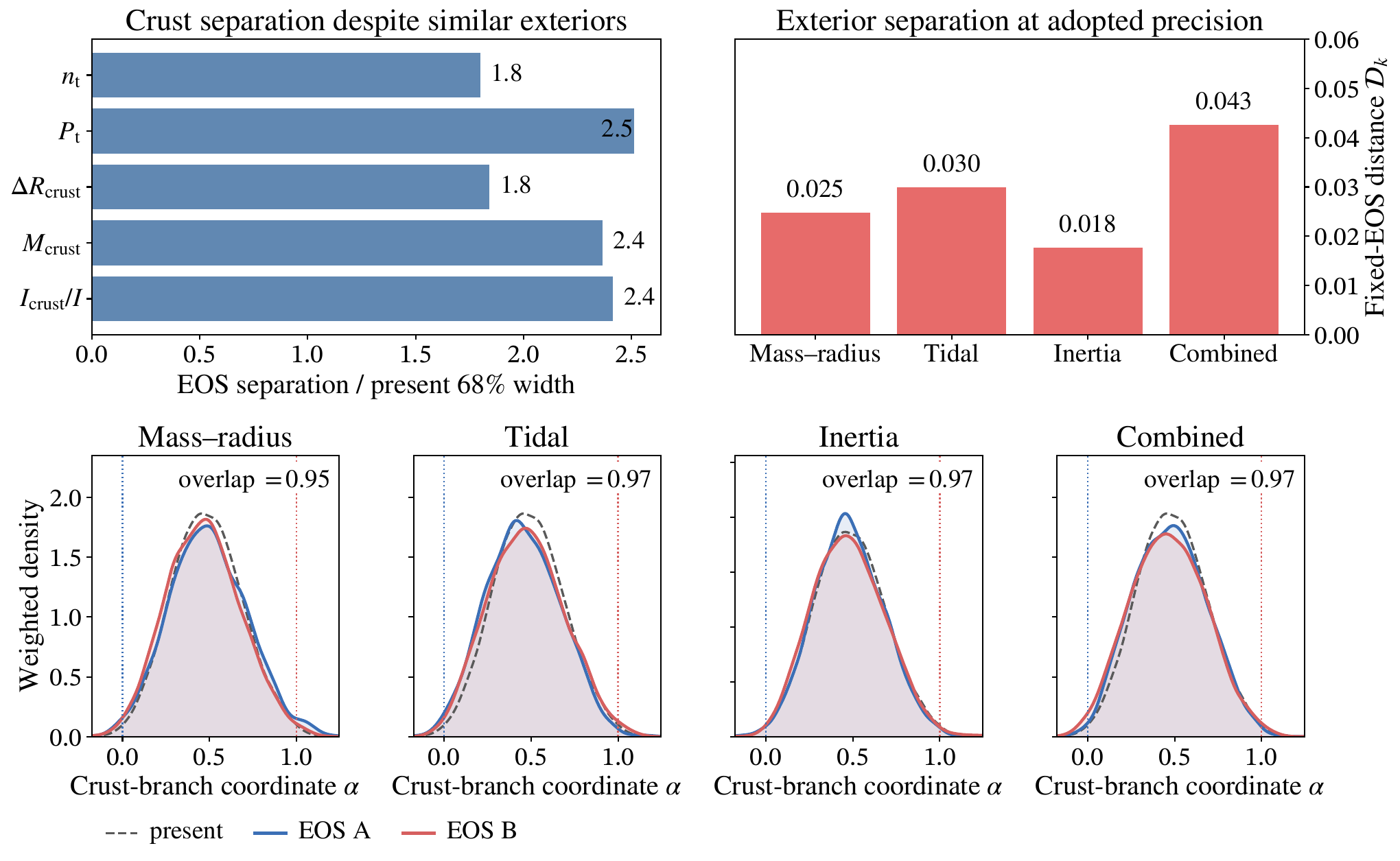}
\caption{Construction and recovery of the crust doppelg\"angers.  The upper
left panel shows the separation between the two EOSs in units of the full
present-data 68\% widths.  The upper right panel gives their distance at the
adopted precision for each measurement.  The lower panels compare the present
posterior with the two matched recoveries along the crust coordinate $\alpha$, scaled so that EOSs $A$ and $B$ lie at
zero and one.  Dotted blue and red lines mark these two EOSs.}
\label{fig:doppel-recovery}
\end{figure*}

We compare EOSs from the full present-data posterior that support all
required global observables.  We quantify their differences in global
observables and crust properties with two dimensionless distances,
\begin{align}
 d_{\rm sel}^2(A,B)={}&
 (\bm O_A-\bm O_B)^T\bm C_{\rm obs}^{-1}
 (\bm O_A-\bm O_B),
 \label{eq:dsel}\\
 d_{\rm crust}^2(A,B)={}&\sum_{X\in\bm y_{\rm crust}}
 \frac{[X_A-X_B]^2}{s_X^2},
 \label{eq:dcrust}
\end{align}
Here $\bm O_A$ and $\bm O_B$ contain the radii, tidal deformabilities,
and maximum masses of the two EOSs.  The covariance matrix $\bm C_{\rm obs}$ accounts for their spread and correlations in the
present posterior, and $T$ denotes transpose.  The quantities $X_A$ and
$X_B$ are their values of crust property $X$, and $s_X$ is its posterior
spread, defined in Appendix~\ref{app:doppel}.  Small $d_{\rm sel}$ and
large $d_{\rm crust}$ identify EOSs with similar global observables but
different crusts.
We select the pair with the largest $d_{\rm crust}$ subject to
$d_{\rm sel}\leq0.5$, obtaining $d_{\rm sel}=0.326$ and
$d_{\rm crust}=9.89$.  Reducing the allowed global differences to $d_{\rm sel}\leq0.25$ still permits a crust
distance of 7.24.  Figure~\ref{fig:doppel-selection-robustness} shows how the
maximum crust separation varies with this limit.

At the masses used in the forecasts, the global observables differ by only
\begin{equation}
\begin{aligned}
 R_A(1.2)-R_B(1.2)&=5.96\times10^{-3}\ {\rm km},\\
 \Lambda_A(1.2)-\Lambda_B(1.2)&=2.89,\\
 I_A(1.338)-I_B(1.338)&=-2.44\times10^{42}\ {\rm g\,cm^2}.
\end{aligned}
\label{eq:doppel-observable-differences}
\end{equation}
For an equal-mass binary, $\widetilde\Lambda=\Lambda$ at the component mass.
The crust properties nevertheless differ substantially
(Table~\ref{tab:doppel-crust-properties}).  Their separation ranges from
1.8 to 2.5 times the present central 68\% width across the five crust quantities.

\begin{table}[t]
\caption{Crust properties of the selected doppelg\"anger EOSs.}
\label{tab:doppel-crust-properties}
\centering
\small
\setlength{\tabcolsep}{3pt}
\begin{ruledtabular}
\begin{tabular}{@{}lrrr@{}}
\toprule
Crust quantity & EOS $A$ & EOS $B$ & $|B-A|/W_{68}^{\rm present}$ \\
\colrule
$n_t$ [fm$^{-3}$] & 0.07207 & 0.09956 & 1.80 \\
$P_t$ [MeV fm$^{-3}$] & 0.2601 & 0.6285 & 2.51 \\
$\Delta R_{\rm crust}(1.4)$ [km] & 0.8898 & 1.1197 & 1.84 \\
$M_{\rm crust}(1.4)$ [$M_\odot$] & 0.01467 & 0.03376 & 2.37 \\
$(I_{\rm crust}/I)(1.4)$ & 0.02157 & 0.04832 & 2.41 \\
\bottomrule
\end{tabular}
\end{ruledtabular}
\tablefoot{The stellar crust quantities are evaluated at $1.4\,\Msun$.
The final column gives the absolute difference between EOSs $B$ and $A$
in units of the full present-data central 68\% width,
$W_{68}^{\rm present}$, for each quantity.}
\end{table}

The measurements used to test this pair are listed in
Table~\ref{tab:doppel-design}.  We express their separation through the
dimensionless distance $\mathcal D$, which measures the difference between
the two EOS predictions relative to the assumed measurement uncertainties,
accounting for correlations between observables:
\begin{equation}
 \mathcal D_k^2(A,B)=
 (\bm\mu_A-\bm\mu_B)^T\bm\Sigma_k^{-1}
 (\bm\mu_A-\bm\mu_B).
 \label{eq:dmeas}
\end{equation}
Here $\bm\mu_A$ and $\bm\mu_B$ are the predicted measurement vectors,
and $\bm\Sigma_k$ is the assumed covariance for measurement type $k$.
At the adopted uncertainties, these differences correspond to distances
$\mathcal D_R=0.025$, $\mathcal D_{\widetilde\Lambda}=0.030$, and
$\mathcal D_I=0.018$.  Combining independent
measurements gives $\mathcal D_{\rm combined}=0.043$.  Reaching
$\mathcal D=1$ would require reducing the mass--radius, tidal, and inertia
uncertainties by factors of 40, 33, and 57, respectively, or by a factor of
23 when all three are combined.  Table~\ref{tab:doppel-precision} also gives
the factors for $\mathcal D=3$.  These distances measure how well the adopted observations separate two
fixed EOSs.  They do not give a detector significance.

For mass and radius, a factor of 40 corresponds to reducing the uncertainty
from 2\% to about 0.05\%, well below the 5\% assumed in a representative
STROBE-X forecast \citep{Ray2019STROBEX}.  Extending the same fractional
tidal precision to $2.0\,\Msun$ raises the separation only to
$\mathcal D=0.43$.  This is optimistic because tidal uncertainties are
expected to increase at higher mass.

\begin{table}[t]
\caption{Precision needed to separate the selected crust doppelg\"angers.}
\label{tab:doppel-precision}
\centering
\small
\setlength{\tabcolsep}{3pt}
\begin{tabularx}{\columnwidth}{@{}lXrrr@{}}
\toprule
Class & Adopted fractional uncertainty & $\mathcal D_0$ & $F_1$ & $F_3$ \\
\midrule
Mass--radius & 2\% in both $M$ and $R$ & 0.0248 & 40.3 & 120.8 \\
Tidal & 10\% in $\widetilde\Lambda$ & 0.0299 & 33.4 & 100.3 \\
Inertia & 10\% in $I$ & 0.0176 & 56.8 & 170.4 \\
Combined & 2\%, 10\%, 10\% & 0.0427 & 23.4 & 70.3 \\
\bottomrule
\end{tabularx}
\tablefoot{$\mathcal D_0$ is their distance at the adopted uncertainty, and
$F_N=N/\mathcal D_0$ is the factor by which every quoted standard deviation
must decrease to reach a fixed-hypothesis distance $N$.  These
detector-independent factors are not detection significances.}
\end{table}

The small measurement distances already indicate that the selected EOSs
cannot be distinguished at the adopted precision.  The posterior recoveries
test whether this ambiguity persists when the remaining EOS parameters are
allowed to vary.  To compare the inferred crusts, we project their properties
onto the scaled line joining $A$ and $B$,
\begin{equation}
 \alpha(\bm y)=
 \frac{\sum_X (y_X-a_X)(b_X-a_X)/w_X^2}
 {\sum_X (b_X-a_X)^2/w_X^2},
 \label{eq:doppel-alpha}
\end{equation}
where $a_X$, $b_X$, and $y_X$ are the values of crust quantity $X$ for
EOSs $A$, $B$, and an inferred EOS, respectively.  The vector $\bm y$
contains the five crust quantities, and $w_X=W_{68}^{\rm present}(X)$
puts them on comparable scales.  Thus $\alpha_A=0$ and $\alpha_B=1$.
The two recovered distributions shown in Figure~\ref{fig:doppel-recovery}
overlap by 0.954--0.975, with Jensen--Shannon divergences at most approximately 0.0035 bits
(Table~\ref{tab:appendix-doppel-recovery}).
The probability assigned to crusts closer to the injected EOS ranges from
0.433 to 0.588.  It changes by at most 0.053 from the present-data value.
The recovered medians remain at $0.447\leq q_{50}(\alpha)\leq0.486$, and neither
endpoint lies within the central 90\% interval: the recovered distributions
concentrate at intermediate values of $\alpha$, rather than near either
injected endpoint.  Combining the radius measurement at $1.2\,\Msun$,
the tidal measurement for the $1.2+1.2\,\Msun$ binary, and the inertia
measurement at $1.338\,\Msun$ does not distinguish the injected crusts.

The pair demonstrates that similar global observables can coexist with
substantially different crusts within the adopted model.  It does not determine
how frequently such pairs occur.  Measurements of radius at percent-level precision, low-mass tidal deformability
at 10\% precision, and total moment of inertia at 10\% precision can remain
compatible with factor-of-two differences in
crust mass and crustal inertia.  Distinguishing these crusts would require
stronger constraints on neutron-rich matter or measurements that probe the
crust more directly.

The crust can affect global stellar properties without being uniquely
constrained by their measurement.  \citet{Fortin2016} showed that
crust--core matching affects stellar radii and crust thicknesses, while
\citet{Gamba2020} found that the assumed crust changes radii inferred from
GW170817 despite its weak effect on the inferred masses and tidal
deformabilities.  Bayesian crust calculations also demonstrate the
constraining role of neutron-skin data and microscopic neutron-matter
calculations \citep{Newton2022Skins}.  The tidal degeneracies identified by
\citet{RaithelMost2023PRD} further illustrate that different EOSs can share
similar global observables.  Our results extend this picture by showing
that, within a unified model, even combined radius, tidal-deformability,
and moment-of-inertia measurements at the assumed precision can leave
substantial differences in crust mass and crustal moment of inertia
unresolved.  Complementary probes may provide information on the crust's
mechanical, superfluid, and thermal properties.  Resonant excitation of the
crust--core interface mode ($i$-mode) during inspiral may trigger a shattering
flare \citep{Tsang2012Shattering,Neill2023RSF}, while magnetar
quasi-periodic oscillations (QPOs), pulsar glitches, and crust cooling probe
other aspects of crust physics
\citep{Israel2005QPO,StrohmayerWatts2006,SamuelssonAndersson2007,Tews2017,
Chamel2013Entrainment,BrownCumming2009}.  Their interpretation requires
additional modelling of composition, magnetic fields, superfluidity,
entrainment, and mode identification.

\section{Conclusions}
\label{sec:conclusion}

A precise determination of global neutron-star properties does not
necessarily imply a precise determination of the crust.  Within our unified
model, present crust constraints are driven mainly by \chEFT{}, with
laboratory data providing additional information and current NICER and
GW170817 observations having a weaker effect.  This pattern persists across
the stellar masses considered.  The inferred precision depends on the
adopted \chEFT{} bands, while the treatment of neutron-rich surface energy
changes the inferred crust properties without appreciably changing the
credible widths at fixed surface prescription.

Future measurements of radius, tidal deformability, and total moment of
inertia mainly improve constraints on crust thickness.  Combining these
measurements gives modest additional gains at matched precision, with some
improvement in transition density but little in transition pressure, crust
mass, or crustal moment of inertia.  In the forecasts considered, stellar
measurements do not recover the precision lost when \chEFT{} is removed.
These findings are conditional on the chosen reference EOS and the adopted
nuclear model.  The doppelg\"anger recoveries further show that substantially
different crusts can remain compatible with nearly identical global
observables under the tested combination of low-mass radius, tidal,
and moment-of-inertia measurements.

The conclusions apply to smooth nucleonic EOSs with a fixed outer crust and
a spherical CLDM.  We expect the main qualitative conclusion, that global
stellar measurements leave substantial uncertainty in crust properties, to
persist when nuclear pasta or alternative nucleonic models are considered.
Progress therefore requires better constraints on low-density neutron-rich
matter and its surface energy, together with observations that probe the
crust more directly.  Resonant shattering flares, magnetar oscillations,
pulsar glitches, and crust cooling offer complementary information beyond
global stellar measurements.

\FloatBarrier
\bibliographystyle{aa_url}
\bibliography{references}

\onecolumn
\begin{appendix}
\begin{multicols}{2}
\raggedcolumns
\nolinenumbers

\section{Unified-model equations}
\label{app:model}

\subsection{Fourth-order nuclear metamodel}

At zero temperature, we describe uniform nucleonic matter by the baryon
density $n=n_n+n_p$, asymmetry $\delta=(n_n-n_p)/n=1-2Y_p$, and
$x=(n-\nsat)/(3\nsat)$.  Here $n_n$ and $n_p$ are neutron and proton
number densities, $Y_p=n_p/n$ is the proton fraction, and $x$ is the
scaled displacement from saturation.  The energy per baryon is
\begin{equation}
 e(n,\delta)=t^{\ast}(n,\delta)+
 \sum_{j=0}^{4}\frac{x^{j}}{j!}
 \left(v_{j}^{\rm is}+v_{j}^{\rm iv}\delta^2\right)
 u_{j}^{(4)}(x).
 \label{eq:metamodel}
\end{equation}
Here $j=0,\ldots,4$ labels the order in the density expansion,
and the superscript $(4)$ denotes truncation at fourth order.
The term $t^{\ast}$ is the kinetic contribution, including the effective-mass
dependence, and $v_{j}^{\rm is}$ and $v_{j}^{\rm iv}$ are the
isoscalar and isovector interaction coefficients.  The empirical parameters
defined in Sect.~\ref{sec:model} fix them algebraically.  The
low-density correction factor is
\begin{equation}
 u_{j}^{(4)}(x)=1-(-3x)^{5-j}\exp[-b(1+3x)].
 \label{eq:regulator}
\end{equation}
This factor ensures that the energy per baryon vanishes at zero density
without changing its value or its first four density derivatives at
saturation.  It follows the low-density correction introduced by
\citet{Margueron2018I}.  Their benchmarks additionally adjust the
high-order coefficients to reproduce the density dependence of reference
EOSs, demonstrating the flexibility of the fourth-order expansion.  These
benchmarks do not assign an uncertainty to terms beyond fourth order.  Because we
do not repeat the inference at another order or with a different low-density
correction, the reported credible intervals do not include this model
dependence.  Cold,
fully equilibrated core matter contains neutrons,
protons, electrons, and muons in beta equilibrium and satisfies charge neutrality.
Accepted EOSs must have positive symmetry energy, mechanical stability,
thermodynamic consistency, and subluminal sound speed throughout the density
range used by the stellar calculation.

\subsection{Compressible liquid-drop crust}

In a spherical Wigner--Seitz cell, the CLDM energy density, minimized over
composition and geometry, is
\begin{equation}
 \begin{split}
 \mathcal{E}_{\rm cell}={}&u\mathcal{E}_{\rm bulk}(n_i,\delta_i)
 +(1-u)\mathcal{E}_n(n_g)+\mathcal{E}_e\\
 &+\mathcal{E}_{\rm surf}+\mathcal{E}_{\rm curv}
 +\mathcal{E}_{\rm Coul},
 \end{split}
 \label{eq:cldm}
\end{equation}
where $u$ is the cluster volume fraction, $n_i$ and $\delta_i$ are
the baryon density and isospin asymmetry of the
cluster interior, respectively, and $n_g$ is the dripped-neutron density.  The first three
terms are the bulk cluster, neutron-gas, and electron energy densities.  The remaining
terms are the surface, curvature, and Coulomb costs of a finite cluster.  We
vary the composition and geometry while conserving baryon number and charge.
The absence of dripped neutrons distinguishes the outer crust, for which
we use the fixed EOS described in Sect.~\ref{sec:model}.

Writing the cluster proton fraction as $Y_{p,i}=(1-\delta_i)/2$, we use the
surface tension \citep{Ravenhall1983Surface}
\begin{equation}
 \sigma_s(Y_{p,i})=\sigma_0
 \frac{2^{p+1}+b_s}{Y_{p,i}^{-p}+b_s+(1-Y_{p,i})^{-p}},
 \label{eq:surface}
\end{equation}
where $\sigma_0$ is the surface tension of symmetric matter and $b_s$
controls its dependence on proton fraction.  The exponent $p$ sets the
neutron-rich extrapolation.  The curvature coefficient is
$\sigma_c=\sigma_s(\sigma_{0,c}/\sigma_0)\alpha_c(\beta-Y_{p,i})$, where
$\sigma_{0,c}$ sets its scale, $\beta$ controls its proton-fraction dependence,
and $\alpha_c=5.5$ is fixed
\citep{Newton2013CLDM}.  For every sampled bulk EOS,
we refit the coefficients
$\{\sigma_0,b_s,\sigma_{0,c},\beta\}$ by least squares to
masses from the 2020 Atomic Mass Evaluation (AME2020)
\citep{Wang2021AME2020}.  The exponent $p$ is
fixed within each calculation.  We use
$p=3$ as the reference and repeat the calculation at $p=2.5$ and $3.5$ to
measure the sensitivity to the surface energy of very neutron-rich nuclei.
The crust--core transition occurs where uniform beta-equilibrated matter
becomes energetically favoured over the optimized nuclear phase.

\section{Priors and likelihoods}
\label{app:inference}

\subsection{Prior and present-data likelihood}
\label{app:present-likelihood}

For data $D$ and model $\mathcal M$, Bayes' theorem gives
\begin{equation}
p(\bm{\theta}|D,\mathcal M)=
 \frac{\mathcal L(D|\bm{\theta},\mathcal M)\pi(\bm{\theta}|\mathcal M)}
 {\mathcal Z(D|\mathcal M)},\qquad
 \mathcal Z=\int\mathcal L\pi\,\dd\bm{\theta}.
 \label{eq:bayes}
\end{equation}
Here $p(\bm\theta|D,\mathcal M)$ is the posterior density, $\pi$ is the
prior density, $\mathcal L$ is the likelihood, and the
evidence $\mathcal Z$ normalizes the posterior.  We use $\mathcal Z$ only for
normalization, not to compare cases that contain different data.  Table~\ref{tab:priors} lists the independent uniform priors on the metamodel
parameters.

\end{multicols}
\begin{appendixtable}
\caption{Continuous uniform priors.}
\label{tab:priors}
\centering
\begin{tabular}{lcl@{\qquad}lcl}
\toprule
Parameter & Prior & Unit & Parameter & Prior & Unit \\
\midrule
$\nsat$ & $[0.150,0.170]$ & fm$^{-3}$ & $\Esat$ & $[-17,-15]$ & MeV \\
$\Ksat$ & $[190,320]$ & MeV & $\Qsat$ & $[-1000,1000]$ & MeV \\
$\Zsat$ & $[-3000,3000]$ & MeV & $\Jsym$ & $[25,40]$ & MeV \\
$\Lsym$ & $[10,100]$ & MeV & $\Ksym$ & $[-400,400]$ & MeV \\
$\Qsym$ & $[-2000,2000]$ & MeV & $\Zsym$ & $[-5000,5000]$ & MeV \\
$m^{\ast}/m$ & $[0.60,0.80]$ & --- & $\Delta m^{\ast}$ & $[0,0.20]$ & --- \\
$b$ & $[1,10]$ & --- & & & \\
\bottomrule
\end{tabular}
\tablefoot{Energy coefficients are in MeV.  The
effective-mass parameters are dimensionless and defined in
Sect.~\ref{sec:model}.}
\end{appendixtable}
\begin{multicols}{2}
\raggedcolumns

The physical term in Eq.~\eqref{eq:factorization} requires a valid unified EOS
that satisfies the conditions in Appendix~\ref{app:model}.  We treat the
observed maximum mass as a lower bound.  EOSs with $\Mmax$ well above $2\,\Msun$ receive
the same weight, while a one-sided Gaussian turn-on of width $0.02\,\Msun$
smoothly suppresses models below it.  This threshold is consistent with the
observed two-solar-mass pulsars \citep{Antoniadis2013,Cromartie2020}.

Table~\ref{tab:nuclear-constraints} summarizes all terms in the present-data
likelihood.  For the Gaussian terms in $\mathcal L_{\rm sat}$ and
$\mathcal L_{\rm lab}$, a quoted value
$\mu\pm\sigma$ contributes a normal density
$\mathcal N[X_{\bm\theta}(n);\mu,\sigma]$, where $\mu$ is the measured
central value, $\sigma$ its standard deviation, and $X_{\bm\theta}(n)$ the
corresponding prediction of the sampled EOS at density $n$.  The adopted saturation constraints and their supporting sources are
listed in Table~\ref{tab:nuclear-constraints}.  The
finite-density set is the subset of the experimental compilation of
\citet{Tsang2024DataCentric} used in our calculation.  The laboratory constraints are treated as independent.

\end{multicols}
\begin{appendixtable}
\caption{Nuclear and astrophysical inputs to the present-data likelihood.}
\label{tab:nuclear-constraints}
\centering
\small
\setlength{\tabcolsep}{4pt}
\begin{tabularx}{\textwidth}{@{}p{0.10\textwidth}>{\raggedright\arraybackslash}p{0.22\textwidth}p{0.09\textwidth}>{\raggedright\arraybackslash}p{0.30\textwidth}>{\raggedright\arraybackslash}X@{}}
\toprule
Likelihood & Physical input & $n$ [fm$^{-3}$] & Adopted constraint or likelihood & Source \\
\midrule
$\mathcal L_{\rm phys}$ & Physical admissibility & Full EOS &
Thermodynamic consistency, stability, positive symmetry energy, and causality &
Appendix~\ref{app:model} \\
\addlinespace
$\mathcal L_{\rm sat}$ & Saturation density $\nsat$ & --- &
$0.155\pm0.005$ fm$^{-3}$ &
\citet{Buchinger1994Radii,Margueron2018I} \\
 & Saturation energy $\Esat$ & --- &
$-16.0\pm0.2$ MeV &
\citet{Chabanat1997SLy,RocaMaza2015Covariance} \\
 & Incompressibility $\Ksat$ & --- &
$230\pm40$ MeV &
\citet{Khan2012Incompressibility} \\
\addlinespace
$\mathcal L_{\rm lab}$ & Dipole polarizability: $S(n)$ & 0.050 &
$15.9\pm1.0$ &
\citet{ZhangChen2015Dipole} \\
 & Isobaric analogue states and isovector skins: $S(n)$ & 0.106 &
$25.5\pm1.1$ &
\citet{Danielewicz2017Skins} \\
 & Isospin diffusion: $S(n)$ & 0.035 &
$10.3\pm1.0$ &
\citet{Tsang2009Symmetry} \\
 & Neutron/proton yields: $S(n)$ & 0.069 &
$16.8\pm1.2$ &
\citet{Morfouace2019Symmetry} \\
 & Pion spectral ratio: $S(n)$ & 0.232 &
$52.0\pm13.0$ &
\citet{Estee2021Pions} \\
 & Pion spectral ratio: $P_{\rm sym}(n)$ & 0.232 &
$10.9\pm8.7$ &
\citet{Estee2021Pions} \\
 & Neutron/proton elliptic flow: $P_{\rm sym}(n)$ & 0.240 &
$12.1\pm8.4$ &
\citet{Cozma2018Elliptic,Russotto2011Flow,Russotto2016ASYEOS} \\
 & Collective flow: $P_{\rm SNM}(n)$ & 0.320 &
$10.1\pm3.0$ &
\citet{Danielewicz2002EOS} \\
\addlinespace
$\mathcal L_{\chi{\rm EFT}}$ & Symmetric and pure-neutron matter energies & 0.02--0.20 &
Uniform acceptance inside the energy bands at the evaluated densities, zero outside. Reference widening $f=0.05$ (Eq.~\eqref{eq:eft-widening}) &
\citet{Drischler2016,Carreau2019Transition,Carreau2019Moment} \\
\addlinespace
$\mathcal L_{\Mmax}$ & Maximum stellar mass & --- &
One-sided Gaussian suppression below $2\,\Msun$, with width $0.02\,\Msun$ &
\citet{Antoniadis2013,Cromartie2020} \\
\addlinespace
$\mathcal L_{\rm NICER}$ & PSR J0030+0451: $M,R$ & --- &
$M=1.34^{+0.15}_{-0.16}\,\Msun$,\par\vspace{3pt}$R=12.71^{+1.14}_{-1.19}$ km &
\citet{Riley2019J0030} \\
\addlinespace[3pt]
 & PSR J0437--4715: $M,R$ & --- &
$M=1.418\pm0.037\,\Msun$,\par\vspace{3pt}$R=11.36^{+0.95}_{-0.63}$ km &
\citet{Choudhury2024J0437} \\
\addlinespace[3pt]
 & PSR J0614--3329: $M,R$ & --- &
$M=1.44^{+0.06}_{-0.07}\,\Msun$,\par\vspace{3pt}$R=10.29^{+1.01}_{-0.86}$ km &
\citet{Mauviard2025J0614} \\
\addlinespace[3pt]
 & PSR J0740+6620: $M,R$ & --- &
$M=2.072^{+0.067}_{-0.066}\,\Msun$,\par\vspace{3pt}$R=12.39^{+1.30}_{-0.98}$ km &
\citet{Riley2021J0740} \\
\addlinespace
$\mathcal L_{\rm GW170817}$ & Binary tidal deformability $\widetilde\Lambda$ and mass ratio $q$ & --- &
Public EOS-insensitive posterior in $(q,\widetilde\Lambda)$, marginalized over $0.7\leq q\leq1$ at $\mathcal M_c=1.186\,\Msun$ &
\citet{Abbott2018GW170817EOS} \\
\bottomrule
\end{tabularx}
\tablefoot{$S(n)$ is the symmetry energy, $L(n)$ is its local density slope,
$P_{\rm sym}(n)$ is the symmetry pressure, and $P_{\rm SNM}(n)$ is the pressure
of symmetric nuclear matter.  Energies are in MeV and pressures in MeV fm$^{-3}$.
For the nuclear Gaussian terms, quoted uncertainties are one standard deviation.
The \chEFT{} widening is applied to the absolute boundary energies.
Sensitivity variants use $f=0$, $0.10$, $0.20$, or omit this term.
NICER values summarize the published marginal 68\% credible intervals for
mass and equatorial radius.  Each NICER source density is
integrated along the predicted mass--radius curve using Eq.~\eqref{eq:nicer},
and the four source terms are multiplied.  The chirp mass $\mathcal M_c$,
mass ratio $q$, and tidal parameter $\widetilde\Lambda$ are defined below.}
\end{appendixtable}
\begin{multicols}{2}
\raggedcolumns

We compare symmetric and pure-neutron matter with the \chEFT{} envelope at
densities from 0.02 to $0.20\,\mathrm{fm}^{-3}$:
\begin{equation}
 \ln\mathcal L_{\chi{\rm EFT}}=
 \begin{cases}
 0,&e_{\rm low}(n_j)\leq e_{\bm\theta}(n_j)\leq e_{\rm high}(n_j)
 \quad\forall j,\\
 -\infty,&\text{otherwise}.
 \end{cases}
 \label{eq:efttophat}
\end{equation}
Here $n_j$ denotes an evaluated density, with $j$ indexing these densities.
The quantity $e_{\bm\theta}$ is the model energy per baryon, and $e_{\rm low}$ and $e_{\rm high}$ are the lower and
upper acceptance limits.  The inequalities must hold for both symmetric and
pure-neutron matter.  We use the bands published with the NSEoS implementation
\citep{Carreau2019Transition,Carreau2019Moment}.  They draw on the chiral two-
and three-nucleon calculations of \citet{Drischler2016}.  Every EOS inside
the envelope has the same likelihood.  In the reference calculation we widen
both edges outwards by a fraction $f=0.05$ of their absolute energies:
\begin{equation}
 e_{\rm low}'=e_{\rm low}-f|e_{\rm low}|,\qquad
 e_{\rm high}'=e_{\rm high}+f|e_{\rm high}|.
\label{eq:eft-widening}
\end{equation}
The primed limits are used in Eq.~\eqref{eq:efttophat}.  Thus the percentage
refers to the boundary energies, not to the original band width.  The
sensitivity calculations use the original \chEFT{} bands ($f=0$),
widenings of $f=0.10$ and $0.20$, or omit the \chEFT{} constraint entirely.

For each NICER source $s$, we evaluate the reconstructed mass--radius
density along the EOS curve,
\begin{equation}
 \mathcal L_s(\bm\theta)\propto\int\dd M\,
 p_s[M,R_{\bm\theta}(M)\mid D_s].
 \label{eq:nicer}
\end{equation}
Here $R_{\bm\theta}(M)$ is the predicted radius, $D_s$ denotes the
observations included in the published source analysis, and $p_s$ is its
reconstructed joint mass--radius posterior density.  This prescription uses
a flat reference prior in mass and radius.  Any radio timing information
included in $D_s$ is retained in $p_s$.  We include
J0030+0451 \citep{Riley2019J0030}, J0437--4715
\citep{Choudhury2024J0437}, J0614--3329 \citep{Mauviard2025J0614}, and
J0740+6620 \citep{Riley2021J0740}.

For GW170817, we use the public EOS-insensitive posterior in the
mass ratio $q=m_2/m_1\leq1$ and the binary tidal parameter
$\widetilde\Lambda$ \citep{Abbott2018GW170817EOS}.  Here $m_1$ and $m_2$
are the source-frame component masses, and
\begin{equation}
 \widetilde\Lambda=\frac{16}{13}
 \frac{(m_1+12m_2)m_1^4\Lambda_1+(m_2+12m_1)m_2^4\Lambda_2}
 {(m_1+m_2)^5}.
 \label{eq:lambdatilde}
\end{equation}
Here $\Lambda_1=\Lambda(m_1)$ and $\Lambda_2=\Lambda(m_2)$ are evaluated
from the sampled EOS.  We marginalize over $q\geq0.7$ at fixed source-frame
chirp mass $\mathcal M_c=(m_1m_2)^{3/5}/(m_1+m_2)^{1/5}=1.186\,\Msun$.  This retains the binary-mass
uncertainty while testing the complete tidal curve rather than one value of
$\Lambda$.

\subsection{Forecast likelihoods and reference EOS}
\label{app:forecast-likelihood}

We choose the reference EOS $T_{\rm REF}$ from the full present-data posterior, near the nuclear-parameter medians.  We require the reference EOS to support the full forecast mass range,
with a small margin above the highest target mass.

For a mass--radius observation centred on
$\bm\mu_{MR}=(M_\star,R_{T_{\rm REF}}(M_\star))$, where $M_\star$ is the
injected stellar mass and $R_{T_{\rm REF}}$ is the reference EOS radius,
the model-curve likelihood is
\begin{equation}
 \mathcal L_{MR}(\bm\theta)=\int\dd M
 \mathcal N_2\!\left(
 \begin{bmatrix}M\\R_{\bm\theta}(M)\end{bmatrix};
 \bm\mu_{MR},\bm\Sigma_{MR}\right),
 \label{eq:futuremr}
\end{equation}
where $\mathcal N_2$ is the bivariate normal density with centre
$\bm\mu_{MR}$ and covariance $\bm\Sigma_{MR}$.  The integral follows the
predicted $R_{\bm\theta}(M)$ curve and therefore marginalizes over the measured
mass rather than evaluating the EOS at one fixed mass.  We use independent
2\% standard deviations in $M$ and $R$.  The tidal forecasts
use a reference equal-mass binary with component mass $m_\star$ and a
5\%, 10\%, or 20\% Gaussian uncertainty on $\widetilde\Lambda$.
We marginalize over a Gaussian mass-ratio distribution peaking at $q=1$,
with Gaussian width $0.03$, truncated to $0.85\leq q\leq1$, at fixed
injected chirp mass.  The
moment-of-inertia likelihood is
\begin{equation}
 \mathcal L_I(\bm\theta)=\mathcal N[I_{\bm\theta}(1.338\,\Msun);
 I_{T_{\rm REF}},f_I I_{T_{\rm REF}}],
 \label{eq:futureI}
\end{equation}
where $I_{\bm\theta}$ is the predicted moment of inertia,
$I_{T_{\rm REF}}=I_{T_{\rm REF}}(1.338\,\Msun)$ is its injected value, and
$f_I=0.05$ or 0.10 is the fractional standard deviation.  Writing
$D_{\rm present}$ and $D_{\rm future}$ for the present and forecast data,
with $\mathcal L_{\rm future}$ denoting the likelihood of the added
measurements, each forecast samples
\begin{equation}
 p(\bm\theta|D_{\rm present},D_{\rm future})\propto
 \mathcal L_{\rm present}\mathcal L_{\rm future}\pi(\bm\theta).
 \label{eq:futureposterior}
\end{equation}

The corresponding no-\chEFT{} posterior is
\begin{equation}
 p_{\neg\chi{\rm EFT}}(\bm\theta|D_{\rm present},D_{\rm future})
 \propto \mathcal L_{\rm present}^{\neg\chi{\rm EFT}}
 \mathcal L_{\rm future}\pi(\bm\theta),
 \label{eq:noeftfutureposterior}
\end{equation}
where the superscript $\neg\chi{\rm EFT}$ denotes omission of the \chEFT{}
factor from the present likelihood.  The reference EOS and future likelihood
$\mathcal L_{\rm future}$ are unchanged.  We repeat four forecasts:
five 2\% radius measurements, a 5\% $\widetilde\Lambda_{1.4+1.4}$ measurement,
the realistic combined programme, and the multi-star combined programme.
Each forecast differs from its \chEFT{}-conditioned counterpart only
through removal of the \chEFT{} term.

\section{Crust-doppelg\"anger definition}
\label{app:doppel}

The observable vector $\bm O$ in Eq.~\eqref{eq:dsel} contains radii at
$1.2$, $1.4$, and $1.8\,\Msun$, tidal deformabilities at $1.2$,
$1.4$, and $1.6\,\Msun$, and the maximum mass $\Mmax$.
Moment of inertia is reserved for an independent test.  The selection
distances are defined in Eqs.~\eqref{eq:dsel} and \eqref{eq:dcrust}.
The scale $s_X=1.4826\,\mathrm{median}|X-\mathrm{median}(X)|$, evaluated
with posterior weights, measures the spread while reducing sensitivity to
extreme values.  The factor 1.4826 gives the standard deviation for a
Gaussian distribution \citep{RousseeuwCroux1993}.  We maximize $d_{\rm crust}$
subject to $d_{\rm sel}\leq0.5$.  Figure~\ref{fig:doppel-selection-robustness}
shows the Pareto frontier (the best compromises between similar global
observables and different crusts) \citep{MarlerArora2004}.

The measurement distance is defined in Eq.~\eqref{eq:dmeas}.
Independent measurements add in $\mathcal D^2$.  For a scalar observable
$O$, the fractional standard deviation needed to reach a distance $N$ is

\begin{equation}
 f_{\rm req}(N)=\frac{|O_A-O_B|}{N\bar O},\qquad
 \bar O=\frac{O_A+O_B}{2}.
 \label{eq:freq}
\end{equation}
Here $O_A$ and $O_B$ are the two predictions.  Reducing all standard
deviations by $F_N=N/\mathcal D_k$ gives the same target distance
(Table~\ref{tab:doppel-precision}).  These distances compare fixed EOSs
and are not detection significances for a particular instrument.

The measurements in Table~\ref{tab:doppel-design} were chosen before the
recoveries.  For radius and tidal deformability, we selected the forecast
with the smallest joint covariance volume of the five scaled crust
quantities.  The inertia precision follows the reference forecast for
J0737--3039A.  Both injected EOSs use the same measurement uncertainties.

\par\addvspace{10pt}\noindent\begin{minipage}{\columnwidth}\makeatletter\def\@captype{table}\makeatother
\caption{Forecast measurements for the crust doppelg\"angers.}
\label{tab:doppel-design}
\centering
\small
\begin{tabularx}{\columnwidth}{@{}lX@{}}
\toprule
Class & Added zero-noise measurement \\
\midrule
Mass--radius & $M,R$ at $1.2\,M_\odot$ with independent 2\% errors \\
Tidal & Equal-mass $1.2+1.2\,M_\odot$ binary with 10\% uncertainty in
$\widetilde\Lambda$ and a Gaussian mass-ratio distribution peaking at $q=1$,
with Gaussian width $0.03$, truncated to $0.85\leq q\leq1$ \\
Inertia & $I(1.338\,M_\odot)$ with 10\% uncertainty \\
Combined & all three measurements above \\
\bottomrule
\end{tabularx}
\tablefoot{We
generate each measurement once from EOS $A$ and once from EOS $B$.  We always
combine it with the full present-data constraints.}
\end{minipage}\par\addvspace{10pt}

We compare the recovered crusts using the coordinate $\alpha$ defined
in Eq.~\eqref{eq:doppel-alpha}.  For each
measurement type, let $p_A(\alpha)$ and $p_B(\alpha)$ be the normalized
posterior densities obtained from data generated by $A$ and $B$, respectively.
We report their posterior-overlap coefficient \citep{PastoreCalcagni2019}
\begin{equation}
 \mathcal O_{AB}=\int\min[p_A(\alpha),p_B(\alpha)]\,\dd\alpha,
 \label{eq:doppel-overlap}
\end{equation}
which runs from zero for disjoint distributions to one for identical
distributions.  We also use the symmetric Jensen--Shannon divergence
$D_{\rm JS}$ \citep{Lin1991}
\begin{equation}
 \begin{split}
 D_{\rm JS}={}&\frac12D_{\rm KL}(p_A\Vert p_{\rm mix})
 +\frac12D_{\rm KL}(p_B\Vert p_{\rm mix}),\\
 p_{\rm mix}={}&\frac{p_A+p_B}{2},
 \end{split}
 \label{eq:doppel-js}
\end{equation}
Here $D_{\rm KL}(p\Vert q)=\int p(\alpha)\log_2[p(\alpha)/q(\alpha)]\,\dd\alpha$
is the Kullback--Leibler divergence of Eq.~\eqref{eq:dkl}, applied to the
normalized densities in $\alpha$, and $p_{\rm mix}$ is their equally
weighted mixture.  The Jensen--Shannon divergence is symmetric and ranges
from zero for identical distributions to one bit for distributions with
no overlap.  We also report the probability on the
side of $\alpha=0.5$ containing the injected EOS: $\alpha<0.5$ for $A$
and $\alpha>0.5$ for $B$.  Recovery requires consistency with the injected
EOS and preference over the other member, rather than narrowing alone.
Table~\ref{tab:appendix-doppel-recovery} lists all eight recoveries.

\end{multicols}
\raggedbottom
\noindent\begin{minipage}{\textwidth}\makeatletter\def\@captype{figure}\makeatother
\centering
\includegraphics[width=\textwidth]{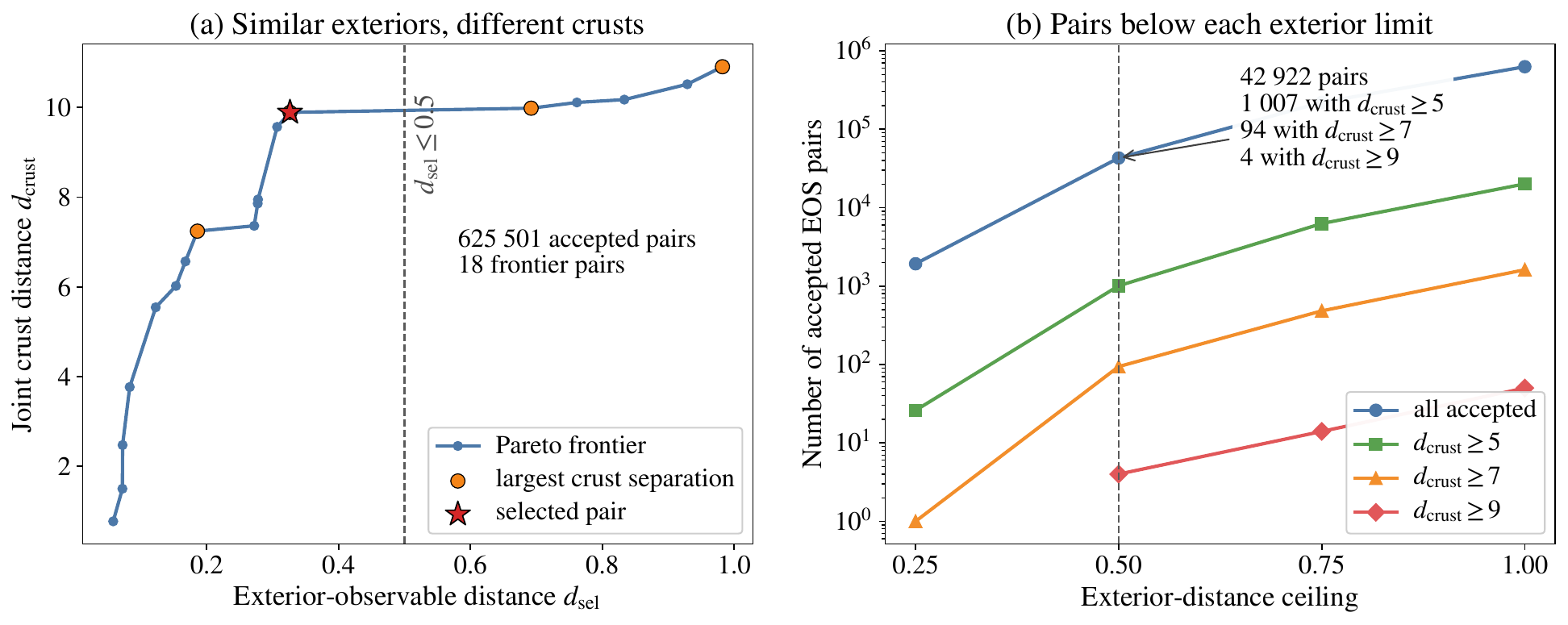}
\caption{Robustness of the crust-doppelg\"anger selection.
\emph{Left:} joint crust distance $d_{\rm crust}$ versus exterior-observable
distance $d_{\rm sel}$ along the Pareto frontier. Orange points mark the
largest crust separations below the tested limits on $d_{\rm sel}$, and
the star marks the selected pair.
\emph{Right:} number of accepted EOS pairs below each exterior-distance
limit, shown for all pairs and for three minimum crust separations.
The vertical dashed lines mark $d_{\rm sel}=0.5$. The annotation in the
right panel gives the pair counts at this limit.}
\label{fig:doppel-selection-robustness}
\end{minipage}
\par\vspace{12pt}
\section{Numerical posterior tables}
\label{app:resulttables}

These tables summarize the posterior constraints and forecast results.

\begingroup

\small
\setlength{\tabcolsep}{3.5pt}

\setlength{\LTcapwidth}{\textwidth}
\begin{longtable}{lccclccc}
\caption{Reference posterior using all present constraints.}\label{tab:appendix-fiducial-posteriors}\\
\toprule
Parameter & median & 68\% & 90\% & Crust quantity & median & 68\% & 90\% \\
\midrule
\endfirsthead
\caption[]{Reference posterior using all present constraints (continued).}\\
\toprule
Parameter & median & 68\% & 90\% & Crust quantity & median & 68\% & 90\% \\
\midrule
\endhead
$n_{\rm sat}$ [fm$^{-3}$] & $0.1601$ & $[0.1561,0.1646]$ & $[0.1537,0.1671]$ & $n_t$ [$\mathrm{fm}^{-3}$] & $0.0811$ & $[0.0742,0.0895]$ & $[0.0704,0.0951]$ \\
$E_{\rm sat}$ [MeV] & $-15.95$ & $[-16.13,-15.76]$ & $[-16.25,-15.65]$ & $P_t$ [MeV $\mathrm{fm}^{-3}$] & $0.396$ & $[0.328,0.475]$ & $[0.291,0.528]$ \\
$K_{\rm sat}$ [MeV] & $231$ & $[215,246]$ & $[206,257]$ & $\Delta R_{\rm crust}(1.4)$ [km] & $1.048$ & $[0.985,1.110]$ & $[0.943,1.151]$ \\
$Q_{\rm sat}$ [MeV] & $-359$ & $[-499,-206]$ & $[-575,-89]$ & $M_{\rm crust}(1.4)$ [$M_\odot$] & $0.0243$ & $[0.0204,0.0285]$ & $[0.0180,0.0314]$ \\
$Z_{\rm sat}$ [MeV] & $674$ & $[-230,1776]$ & $[-573,2436]$ & $(I_{\rm crust}/I)(1.4)$ & $0.0355$ & $[0.0301,0.0412]$ & $[0.0267,0.0452]$ \\
$J_{\rm sym}$ [MeV] & $30.69$ & $[29.70,31.79]$ & $[29.24,32.43]$ &  &  &  &  \\
$L_{\rm sym}$ [MeV] & $47.04$ & $[38.37,54.75]$ & $[33.39,58.61]$ &  &  &  &  \\
$K_{\rm sym}$ [MeV] & $-85$ & $[-137,-32]$ & $[-171,6]$ &  &  &  &  \\
$Q_{\rm sym}$ [MeV] & $965$ & $[613,1327]$ & $[397,1565]$ &  &  &  &  \\
$Z_{\rm sym}$ [MeV] & $-2413$ & $[-4042,-406]$ & $[-4635,879]$ &  &  &  &  \\
$m^\ast/m$ & $0.69$ & $[0.63,0.77]$ & $[0.61,0.79]$ &  &  &  &  \\
$\Delta m^\ast$ & $0.11$ & $[0.04,0.17]$ & $[0.01,0.19]$ &  &  &  &  \\
$b$ & $4.31$ & $[1.89,7.76]$ & $[1.34,9.32]$ &  &  &  &  \\
\bottomrule
\end{longtable}
\nopagebreak

\tablefoot{The left block gives
the thirteen empirical parameters.  The effective-mass parameters and $b$ are dimensionless.  The
right block gives the five crust targets, with mass-labelled quantities at
$1.4\,M_\odot$.  Entries are the median and central 68\% and 90\% intervals.}

\endgroup

\begingroup

\small

\setlength{\LTcapwidth}{\textwidth}
\begin{longtable}{lcccccc}
\caption{Sensitivity to present constraints.}\label{tab:appendix-present-effects}\\
\toprule
Case & $n_t$ & $P_t$ & $\Delta R_{\rm crust}$ & $M_{\rm crust}$
& $I_{\rm crust}/I$ & $\max|\Delta\widetilde X|/W_{68}^{\rm present}$ \\
\midrule
\endfirsthead
\caption[]{Sensitivity to present constraints (continued).}\\
\toprule
Case & $n_t$ & $P_t$ & $\Delta R_{\rm crust}$ & $M_{\rm crust}$
& $I_{\rm crust}/I$ & $\max|\Delta\widetilde X|/W_{68}^{\rm present}$ \\
\midrule
\endhead
physical only & 2.29 & 2.39 & 3.28 & 3.04 & 3.19 & 1.78 \\
nuclear only & 0.84 & 0.86 & 1.11 & 1.02 & 1.01 & 0.15 \\
astrophysical only & 2.73 & 3.54 & 2.91 & 3.65 & 3.78 & 0.98 \\
all present & 1.00 & 1.00 & 1.00 & 1.00 & 1.00 & 0.00 \\
no \chEFT{} & 2.34 & 2.91 & 2.08 & 2.89 & 2.96 & 0.86 \\
no laboratory & 1.27 & 1.24 & 1.11 & 1.20 & 1.19 & 0.32 \\
no NICER & 0.83 & 0.88 & 1.05 & 0.99 & 1.00 & 0.13 \\
no GW170817 & 0.98 & 0.99 & 1.05 & 1.03 & 1.04 & 0.07 \\
\chEFT{} original & 0.48 & 0.42 & 0.78 & 0.54 & 0.53 & 0.25 \\
\chEFT{} 10\% & 1.28 & 1.34 & 1.22 & 1.37 & 1.38 & 0.09 \\
\chEFT{} 20\% & 1.69 & 1.96 & 1.49 & 1.93 & 1.97 & 0.21 \\
surface $p=2.5$ & 1.02 & 0.97 & 1.04 & 1.03 & 1.04 & 0.52 \\
surface $p=3.5$ & 0.97 & 1.03 & 1.01 & 1.06 & 1.06 & 0.32 \\
\bottomrule
\end{longtable}
\nopagebreak

\tablefoot{The five central columns give
$W_{68}^{\rm case}/W_{68}^{\rm present}$.  Values above one are broader than
the reference posterior.  The last column gives the largest absolute normalized median shift,
$\max_X |q_{50}^{\rm case}(X)-q_{50}^{\rm present}(X)|/W_{68}^{\rm present}(X)$,
where $\Delta\widetilde X=q_{50}^{\rm case}(X)-q_{50}^{\rm present}(X)$.}

\endgroup

\begingroup

\setlength{\LTcapwidth}{\textwidth}
\begin{longtable}{cccc}
\caption{Mass dependence of the reference crust posterior.}\label{tab:appendix-mass-dependent}\\
\toprule
$M/M_\odot$ & $\Delta R_{\rm crust}$ [km] & $M_{\rm crust}$ [$M_\odot$]
& $I_{\rm crust}/I$ \\
\midrule
\endfirsthead
\caption[]{Mass dependence of the reference crust posterior (continued).}\\
\toprule
$M/M_\odot$ & $\Delta R_{\rm crust}$ [km] & $M_{\rm crust}$ [$M_\odot$]
& $I_{\rm crust}/I$ \\
\midrule
\endhead
1.0 & $1.573\;[1.492,1.654]$ & $0.0320\;[0.0273,0.0369]$ & $0.0641\;[0.0555,0.0733]$ \\
1.2 & $1.275\;[1.204,1.347]$ & $0.0280\;[0.0237,0.0326]$ & $0.0472\;[0.0403,0.0544]$ \\
1.4 & $1.048\;[0.985,1.110]$ & $0.0243\;[0.0204,0.0285]$ & $0.0355\;[0.0301,0.0412]$ \\
1.6 & $0.864\;[0.809,0.920]$ & $0.0209\;[0.0174,0.0247]$ & $0.0269\;[0.0226,0.0314]$ \\
1.8 & $0.708\;[0.659,0.759]$ & $0.0176\;[0.0145,0.0209]$ & $0.0203\;[0.0169,0.0240]$ \\
2.0 & $0.573\;[0.525,0.619]$ & $0.0143\;[0.0117,0.0172]$ & $0.0151\;[0.0124,0.0180]$ \\
\bottomrule
\end{longtable}
\nopagebreak

\tablefoot{Each cell
reports the median followed by the central 68\% interval $[q_{16},q_{84}]$.}

\endgroup

\begingroup

\small
\setlength{\tabcolsep}{4pt}

\setlength{\LTcapwidth}{\textwidth}
\begin{longtable}{lccccccc}
\caption{Effect of future measurements.}\label{tab:appendix-future-effects}\\
\toprule
Case & $n_t$ & $P_t$ & $\Delta R_{\rm crust}$ & $M_{\rm crust}$
& $I_{\rm crust}/I$ & $G_{\rm geom}$ & $\max D_{\rm KL}$ [bits] \\
\midrule
\endfirsthead
\caption[]{Effect of future measurements (continued).}\\
\toprule
Case & $n_t$ & $P_t$ & $\Delta R_{\rm crust}$ & $M_{\rm crust}$
& $I_{\rm crust}/I$ & $G_{\rm geom}$ & $\max D_{\rm KL}$ [bits] \\
\midrule
\endhead
$R(1.2)$, 2\% & 1.05 & 1.02 & 1.09 & 0.96 & 0.96 & 1.01 & 0.027 \\
$R(1.4)$, 2\% & 1.01 & 1.00 & 1.12 & 0.98 & 0.98 & 1.01 & 0.029 \\
$R(1.6)$, 2\% & 0.98 & 0.98 & 1.12 & 0.97 & 0.97 & 1.00 & 0.035 \\
$R(1.8)$, 2\% & 0.97 & 0.96 & 1.11 & 0.99 & 0.98 & 1.00 & 0.065 \\
five $R(M)$, 2\% & 1.05 & 1.04 & 1.22 & 1.03 & 1.01 & 1.07 & 0.147 \\
$\widetilde\Lambda_{1.2+1.2}$, 10\% & 1.05 & 1.03 & 1.15 & 1.00 & 1.00 & 1.04 & 0.047 \\
$\widetilde\Lambda_{1.4+1.4}$, 10\% & 0.97 & 0.98 & 1.17 & 0.99 & 0.99 & 1.02 & 0.055 \\
$\widetilde\Lambda_{1.6+1.6}$, 10\% & 0.93 & 0.91 & 1.21 & 0.98 & 0.97 & 0.99 & 0.145 \\
$\widetilde\Lambda_{1.4+1.4}$, 5\% & 0.97 & 0.98 & 1.27 & 1.01 & 1.01 & 1.04 & 0.124 \\
$\widetilde\Lambda_{1.4+1.4}$, 20\% & 1.02 & 0.99 & 1.06 & 0.99 & 0.99 & 1.01 & 0.017 \\
$I(1.338)$, 10\% & 1.02 & 0.98 & 0.99 & 0.97 & 0.96 & 0.98 & 0.006 \\
$I(1.338)$, 5\% & 1.01 & 1.02 & 1.09 & 1.00 & 0.99 & 1.02 & 0.021 \\
realistic combined & 0.94 & 0.95 & 1.21 & 1.00 & 0.98 & 1.01 & 0.115 \\
multi-star combined & 1.11 & 1.01 & 1.22 & 1.01 & 0.98 & 1.06 & 0.162 \\
\bottomrule
\end{longtable}
\nopagebreak

\tablefoot{The five crust columns give
$G_X=W_{68}^{\rm present}/W_{68}^{\rm present+future}$.  Values above one indicate
narrower intervals.  $G_{\rm geom}$ is their geometric mean, and the final
column is the largest marginalized information gain for that measurement.}

\endgroup

\begingroup

\small
\setlength{\tabcolsep}{5pt}

\setlength{\LTcapwidth}{\textwidth}
\begin{longtable}{lrrrrrr}
\caption{Recovery of the selected crust doppelg\"angers.}\label{tab:appendix-doppel-recovery}\\
\toprule
Measurement & $\mathcal D$ & $\mathcal O_{AB}$ & $D_{\rm JS}$ [bits]
& $P_A(\alpha<0.5)$ & $P_B(\alpha>0.5)$ & $\max|\Delta P|$ \\
\midrule
\endfirsthead
\caption[]{Recovery of the selected crust doppelg\"angers (continued).}\\
\toprule
Measurement & $\mathcal D$ & $\mathcal O_{AB}$ & $D_{\rm JS}$ [bits]
& $P_A(\alpha<0.5)$ & $P_B(\alpha>0.5)$ & $\max|\Delta P|$ \\
\midrule
\endhead
mass--radius & 0.0248 & 0.954 & 0.00351 & 0.524 & 0.433 & 0.032 \\
tidal & 0.0299 & 0.968 & 0.00233 & 0.588 & 0.442 & 0.053 \\
inertia & 0.0176 & 0.967 & 0.00165 & 0.560 & 0.443 & 0.024 \\
combined & 0.0427 & 0.975 & 0.00194 & 0.553 & 0.434 & 0.030 \\
\bottomrule
\end{longtable}
\nopagebreak

\tablefoot{The crust properties of the two EOSs are listed in
Table~\ref{tab:doppel-crust-properties}.  $\mathcal D$ is the
fixed-EOS distance at the adopted measurement precision, $\mathcal O_{AB}$ is
the overlap of the two recovered distributions in $\alpha$, $D_{\rm JS}$ is
their Jensen--Shannon divergence, and $P_A$ ($P_B$) is the probability assigned
to the correct side of $\alpha=0.5$ for data generated from $A$ ($B$).
The final column is the larger absolute change in this probability relative
to the present-data posterior.}

\endgroup

\end{appendix}

\end{document}